\documentclass[12pt, preprint]{aastex}

\usepackage{natbib}

\newcommand{\nuin}{\nu_{\mathrm{in}}}
\newcommand{\nuni}{\nu_{\mathrm{ni}}}

\newcommand{\ca}{c_{\mathrm{A}}}
\newcommand{\rhon}{\rho_{\rm n}}
\newcommand{\rhoi}{\rho_{\rm i}}

\newcommand{\der}{{\rm d}}
\newcommand{\pd}{\partial}

\newcommand{\ain}{\alpha_{\rm in}}
\newcommand{\gammai}{\Gamma_{\rm i}}
\newcommand{\gamman}{\Gamma_{\rm n}}
\newcommand{\tgammai}{\tilde{\Gamma}_{\rm i}}
\newcommand{\tgamman}{\tilde{\Gamma}_{\rm n}}

\begin{document}

\title{Alfv\'en waves in a partially ionized two-fluid plasma}

\shorttitle{Alfv\'en waves in a two-fluid plasma}

\author{R. Soler$^{1}$, M. Carbonell$^{2}$,  J. L. Ballester$^{1}$, \& J. Terradas$^{1}$}
\affil{$^{1}$Departament de  F\'{\i}sica,  Universitat de les Illes Balears, E-07122 Palma de Mallorca, Spain}
\affil{$^{2}$Departament de Matem\`{a}tiques i Inform\`{a}tica, Universitat de les Illes Balears, E-07122 Palma de Mallorca, Spain}

\email{roberto.soler@uib.es, marc.carbonell@uib.es, joseluis.ballester@uib.es, jaume.terradas@uib.es}    

\begin{abstract}

Alfv\'en waves are a particular class of magnetohydrodynamic waves relevant in many astrophysical and laboratory plasmas. In partially ionized plasmas the dynamics of Alfv\'en waves is affected by the interaction between ionized and neutral species. Here we study Alfv\'en waves in a partially ionized plasma from the theoretical point of view using the two-fluid description. We consider that the plasma is composed of an ion-electron fluid and a neutral fluid, which interact by means of particle collisions. To keep our investigation as general as possible we take the neutral-ion collision frequency and the ionization degree as free parameters. First, we perform a normal mode analysis. We find the modification  due to neutral-ion collisions of the wave frequencies and study the temporal and spatial attenuation of the waves. In addition, we discuss the presence of cut-off values of the wavelength that constrain the existence of oscillatory standing waves in weakly ionized plasmas. Later, we go beyond the normal mode approach and solve the initial-value problem in order to study the time-dependent evolution of the wave perturbations in the two fluids. An application to Alfv\'en waves in the low solar atmospheric plasma is performed and the implication of partial ionization for the energy flux is discussed.

\end{abstract}


\section{Introduction}

Alfv\'en waves are a particular class of magnetohydrodynamic (MHD) waves driven by magnetic tension \citep{alfven1942}. In a uniform and infinite plasma  the motions of Alfv\'en waves are incompressible and polarized perpendicularly to the direction of the magnetic field \citep[see, e.g.,][]{hasegawa1982,cramer2001,goossens2003}.  Alfv\'en waves are found in both laboratory and astrophysical plasmas \citep[see review by][]{gekelman11}.

Since the pioneering  works by, e.g., \citet{piddington1956} and \citet{kulsrud1969} it is known that partial ionization of the plasma affects the dynamics of Alfv\'en waves. The feature most extensively investigated in the literature is the wave damping due to collisions between ions and neutrals, although other effects as, e.g., the existence of cut-off values of the wavelength are also an important consequence of partial ionization \citep{kulsrud1969}.

Most of the works that studied Alfv\'en waves in partially ionized plasmas adopted the so-called single-fluid approximation \citep[see, e.g.,][]{brag}. The single-fluid approximation assumes a strong coupling between ions and neutrals. For an MHD wave, this condition means that the wave frequency has to be  much lower than the frequency at which ion and neutrals collide. In other words, it is necessary that in one period there are enough collisions for ions and neutrals to behave as one fluid. This restriction is fulfilled in, e.g., the partially ionized solar plasma and so the single-fluid approximation is usually adopted in that case \citep[see, e.g.,][among others]{depontieu2001,khodachenko2004,forteza2007,soler2009PI}. 

An alternative approach is the multi-fluid theory \citep[see, e.g.,][]{zaqarashvili2011a}, which considers the various species in the plasma as separate fluids. In the multi-fluid description no restriction is imposed on the relative values of the wave frequency and the collision frequency, although the mathematical treatment is usually more complicated than in the single-fluid case. However the multi-fluid theory has the advantage that it is more general than the single-fluid approximation and can be used regardless the value of the collision frequency. Hence the multi-fluid theory is adequate to study MHD waves in those situations where the single-fluid approximation does not apply. This may be the case of molecular clouds \citep[see, e.g.,][]{pudritz1990,balsara1996,mouschovias2011}.  A particular form of the multi-fluid theory is the two-fluid theory in which ions and electrons are considered together as an ion-electron fluid,  while neutrals form another fluid that interacts with the ion-electron fluid by means of collisions. For the investigation of MHD waves this approach was followed by, e.g., \citet{kumar2003,zaqarashvili2011a,mouschovias2011,soler2012}.

Despite the existing literature on this topic (see the references in the above paragraphs), the purpose of the present article is to revisit the theoretical investigation of Alfv\'en waves in partially ionized plasmas using the two-fluid theory. Our reasons for tackling this task are the following. 

First of all, the existing papers in the literature often focus on very specific situations as, e.g., the interstellar medium \citep[e.g.,][]{kulsrud1969}, molecular clouds \citep[see, e.g.,][]{pudritz1990,balsara1996,mouschovias2011}, and solar plasmas \citep[][]{kumar2003,zaqarashvili2011a,soler2012} among other cases. Here our aim is to keep the  investigation as general as possible. To do so we take the neutral-ion collision frequency and the plasma ionization degree as free parameters. This makes the results of the present article to be widely applicable.  We put emphasis on the mathematical transparency and on the finding of approximate analytic solutions. 

In addition, there has been recently some confusion about the existence of cut-off wavelengths for Alfv\'en waves in a partially ionized  plasma \citep[see][]{zaqarashvili2011a,zaqarashvili2012}.  While \citet{zaqarashvili2012} have shown that the presence of cut-offs in the single-fluid approximation is a mathematical artifact, the existence of cut-offs in the two-fluid case is a real physical phenomenon \citep[e.g.,][]{kulsrud1969,pudritz1990,kamaya1998,mouschovias2011}. An important goal of the present article is to stress the existence of physical cut-offs   for Alfv\'en waves in a two-fluid plasma.

Finally, unlike previous works that are restricted to the normal mode analysis, here we combine results of normal modes with the solution of the initial-value problem. This procedure allows us to investigate how strong is the coupling between the perturbations in the ionized fluid and the neutral fluid depending on the relative values of the wave frequency and the neutral-ion collision frequency.

This paper is organized as follows. Section~\ref{sec:equi} contains the description of the equilibrium configuration and the basic equations of the two-fluid theory. Alfv\'en waves are investigated following a normal mode analysis in Section~\ref{sec:normal}, while the initial-value problem is solved in Section~\ref{sec:initial}. Later, Section~\ref{sec:app} contains an application to the solar atmospheric plasma and Section~\ref{sec:energy} discusses the implications of partial ionization for the energy flux of Alfv\'en waves. Finally the conclusions of this work are given in Section~\ref{sec:con}. 

\section{Equilibrium and basic equations}
\label{sec:equi}

We consider a partially ionized medium composed of ions, electrons, and neutrals. We use the two-fluid theory in which ions and electrons are considered together as an ion-electron fluid, i.e., the ionized fluid, while neutrals form another fluid that interacts with the ionized fluid by means of collisions \citep[see, e.g.,][]{zaqarashvili2011a, soler2012}. In all the following expressions, the subscripts `i' and `n' refer to the ionized fluid and the neutral fluid, respectively.

 The equilibrium is made of a uniform and unbounded partially ionized plasma. We use Cartesian coordinates. The equilibrium magnetic field is straight and constant along the  $z$-direction, namely ${\bf B}=B\, \hat{z}$. We ignore the effect of gravity. We also assume that the equilibrium is static so that there are no equilibrium flows. The governing equations for the various species composing the plasma can be found in, e.g., \citet{zaqarashvili2011a}. Here we restrict ourselves to the study of linear perturbations superimposed on the equilibrium state. Hence, the governing equations are linearized. The resulting equations are
\begin{eqnarray}
 \rhoi \frac{\pd {\bf v}_{\rm i}}{\partial t} &=& - \nabla p_{\rm i} + \frac{1}{\mu}\left( \nabla \times {\bf b} \right) \times {\bf B} -\ain \left( {\bf v}_{\rm i} - {\bf v}_{\rm n}\right),  \label{eq:momlinion}\\ 
\rhon \frac{\pd {\bf v}_{\rm n}}{\partial t} &=& - \nabla p_{\rm n} -\ain \left( {\bf v}_{\rm n} - {\bf v}_{\rm i}\right), \label{eq:momlinneu}\\
\frac{\pd {\bf b}}{\partial t} &=& \nabla \times \left( {\bf v}_{\rm i} \times {\bf B} \right), \label{eq:inductionlin} \\
\frac{\partial p_{\rm i}}{\partial t}  &=&   - \gamma P_{\rm i} \nabla \cdot  {\bf v}_{\rm i},  \label{eq:presslinion} \\
\frac{\partial p_{\rm n}}{\partial t}  &=& - \gamma P_{\rm n} \nabla \cdot  {\bf v}_{\rm n},  \label{eq:presslin} 
\end{eqnarray}
where ${\bf v}_{\rm i}$, $p_{\rm i}$, $P_{\rm i}$ and $\rhoi$ are the velocity perturbation, pressure perturbation, equilibrium pressure, and equilibrium density of the ionized fluid, ${\bf v}_{\rm n}$, $p_{\rm n}$, $P_{\rm n}$, and $\rhon$ are the respective quantities but for the neutral fluid, ${\bf b}$ is the magnetic field perturbation, $\mu$ is the magnetic permeability, $\gamma$ is the adiabatic index, and $\ain$ is the ion-neutral friction coefficient. In the specific case of a hydrogen plasma, the expression of $\ain$ is given by \citet{brag}, namely
\begin{equation}
\ain = \frac{1}{2} \frac{\rhoi \rhon}{m_{\rm n}}\sqrt{\frac{16 k_{\rm B} T}{\pi m_{\rm i}}}\sigma_{\rm in}, \label{eq:ainhydrogen}
\end{equation}
where $m_{\rm i}$ and $m_{\rm n}$ are the ion and neutral masses, respectively ($m_{\rm i}\approx m_{\rm n}$ for hydrogen), $k_{\rm B}$ is Boltzmann's constant, $T$ is the plasma temperature, and $\sigma_{\rm in}$ is the collision cross section. In the following analysis we do not use this expression of $\ain$ since we take $\ain$ as a free parameter. We do so to conveniently control the strength of the ion-neutral friction force. Equation~(\ref{eq:ainhydrogen}) is used in the application to solar plasmas done in Section~\ref{sec:app}.

We perform a Fourier analysis of the perturbations in space. In linear theory, an arbitrary perturbation can be represented by the superposition of Fourier components, so that we can restrict ourselves to study particular Fourier components. Therefore, the spatial dependence of perturbations is put proportional to $\exp \left(i k_x x + ik_y y + i k_z z \right)$, where $k_x$, $k_y$, and $k_z$ are the components of the wavenumber in the $x$-, $y$-, and $z$-directions, respectively. In a uniform and infinite plasma Alfv\'en waves are the only MHD modes that propagate vorticity perturbations \citep[see, e.g.,][]{cramer2001,goossens2003}. In addition, Alfv\'en waves are incompressible and their motions are confined to perpendicular planes to the magnetic field, i.e.,  $v_{{\rm i}, z} = v_{{\rm n}, z} = 0$.  Therefore, an appropriate quantity to describe Alfv\'en waves is the vorticity component along the magnetic field direction. By working with vorticity perturbations we are able to decouple Alfv\'en waves from magnetoacoustic waves. We define $\gammai$ and $\gamman$ as the $z$-components of vorticity of the ionized fluid and the neutral fluid, respectively,
\begin{eqnarray}
 \gammai &=& \left( \nabla \times {\bf v}_{\rm i}  \right) \cdot \hat{z} = i k_x v_{{\rm i}, y} - i k_y v_{{\rm i}, x}, \label{eq:defgi} \\
 \gamman &=& \left( \nabla \times {\bf v}_{\rm n}  \right) \cdot \hat{z} = i k_x v_{{\rm n},y} - i k_y v_{{\rm n},x}.\label{eq:defgn}
\end{eqnarray}
Note that in the reference frame in which $k_y=0$, $\gammai$ and $\gamman$ are proportional to $v_{{\rm i}, y}$ and $v_{{\rm n}, y}$, respectively. We combine Equations~(\ref{eq:momlinion})--(\ref{eq:presslin}) and after  some algebraic manipulations we obtain the two following equations involving $\gammai$ and $\gamman$ only, namely
\begin{eqnarray}
 \rhoi \frac{\pd^2 \gammai}{\pd t^2} + \ain \frac{\pd \gammai}{\pd t} + \rhoi k_z^2 \ca^2 \gammai &=& \ain \frac{\pd \gamman}{\pd t}, \label{eq:vorti} \\
\rhon \frac{\pd \gamman}{\pd t} + \ain \gamman &=& \ain \gammai, \label{eq:vortn}
\end{eqnarray}
where $\ca = B/\sqrt{\mu \rhoi}$ is the Alfv\'en velocity. Note that the Alfv\'en velocity is here defined using the density of the ionized fluid only. Equations~(\ref{eq:vorti}) and (\ref{eq:vortn}) are the governing equations for linear vorticity perturbations and, therefore, they are the governing equations of Alfv\'en waves. 

For the subsequent analysis we define the ionization fraction, $\chi$, the ion-neutral collision frequency, $\nu_{\rm in}$, and the neutral-ion collision frequency, $\nu_{\rm ni}$, as follows
\begin{equation}
\chi = \frac{\rhon}{\rhoi}, \qquad \nu_{\rm in} = \frac{\ain}{\rhoi}, \qquad \nu_{\rm ni} = \frac{\ain}{\rhon}.
\end{equation}
Since the collision frequencies are related by $\rhoi \nu_{\rm in} = \rhon \nu_{\rm ni}$, we use $\nu_{\rm ni}$ in all the following expressions for simplicity. Note that when $\rhoi \neq \rhon$, $\nu_{\rm in} \neq \nu_{\rm ni}$, meaning that the ion-neutral and neutral-ion collision frequencies are different \citep[see a discussion on this issue in][]{zaqarashvili2011helium}.

\section{Normal Mode Analysis}
\label{sec:normal}

Here we perform a normal mode analysis. The temporal dependence of the perturbations is put proportional to $\exp\left( -i\omega t \right)$, where $\omega$ is the angular frequency. From Equation~(\ref{eq:vortn}) we express $\gamman$ in terms of $\gammai$ and insert the expression in Equation~(\ref{eq:vorti}). We arrive at an equation involving $\gammai$ only, namely
\begin{equation}
\mathcal{D}\left( \omega, k_z \right) \gammai = 0, \label{eq:gammai0}
\end{equation}
with
\begin{equation}
 \mathcal{D}\left( \omega, k_z \right) = \omega^3 + i \left( 1+\chi \right) \nuni \omega^2 - k_z^2 c_{\rm A}^2 \omega - i \nuni k_z^2 c_{\rm A}^2. \label{eq:relalfpre}
\end{equation}
For $\gammai\neq 0$, the solutions to Equation~(\ref{eq:gammai0}) must satisfy $\mathcal{D}\left( \omega, k_z \right) = 0$, i.e.,
\begin{equation}
 \omega^3 + i \left( 1+\chi \right) \nuni \omega^2 - k_z^2 c_{\rm A}^2 \omega - i \nuni k_z^2 c_{\rm A}^2=0. \label{eq:relalf}
\end{equation}
Equation~(\ref{eq:relalf}) is the dispersion relation of Alfv\'en waves. Although with different notations, Equation~(\ref{eq:relalf}) is equivalent to the dispersion relations previously found by, e.g., \citet{piddington1956,kulsrud1969,pudritz1990,martin1997,kamaya1998,kumar2003,zaqarashvili2011a,mouschovias2011}. In the absence of collisions, $\nuni=0$ and Equation~(\ref{eq:relalf}) becomes
\begin{equation}
\omega \left( \omega^2-k_z^2\ca^2 \right) = 0. \label{eq:relalfideal}
\end{equation}
From Equation~(\ref{eq:relalfideal}) we get the classic dispersion relation of Alfv\'en waves in an ideal plasma, namely $ \omega^2=k_z^2\ca^2 $, and an additional mode with $\omega=0$. The general situation $\nuni\neq 0$ is investigated next.

\subsection{Standing waves}

We focus first on standing waves. Hence we assume a real wavenumber, $k_z$, and solve the dispersion relation (Equation~(\ref{eq:relalf})) to obtain the complex frequency, $\omega= \omega_{\rm R} + i \omega_{\rm I}$, with $\omega_{\rm R}$ and $\omega_{\rm I}$ the real and imaginary parts of $\omega$, respectively. Since $\omega$ is complex the amplitude of perturbations is multiplied by the factor $\exp(\omega_{\rm I} t)$, with $\omega_{\rm I} < 0$. Therefore the perturbations are damped in time. 

Equation~(\ref{eq:relalf}) is a cubic equation so it has three solutions. Unfortunately the exact analytic solution to Equation~(\ref{eq:relalf}) is too complicated to shed any light on the physics. However we can investigate the nature of the solutions using the concept of the polynomial discriminant. We perform the change of variable $\omega = -i s$, so that Equation~(\ref{eq:relalf}) becomes
\begin{equation}
 s^3 +  \left( 1+\chi \right) \nuni s^2 + k_z^2 c_{\rm A}^2 s + \nuni k_z^2 c_{\rm A}^2 = 0. \label{eq:relalf1}
\end{equation}
Equation~(\ref{eq:relalf1}) is a cubic equation and all its coefficients are real. From Equation~(\ref{eq:relalf1}) we  compute the discriminant, $\Lambda$, namely \citep[see, e.g.,][]{cohen2000}
\begin{eqnarray}
 \Lambda = &-& k_z^2 c_{\rm A}^2 \left[ 4\left(1+\chi \right)^3 \nuni^4 \right. \nonumber \\
  &-& \left. \left(  \chi^2 + 20\chi -8\right)\nuni^2 k_z^2 c_{\rm A}^2 +4 k_z^4 c_{\rm A}^4\right],
  \label{eq:relalf2u}
\end{eqnarray}
The discriminant, $\Lambda$, is defined so that (i) Equation~(\ref{eq:relalf1}) has one real zero and two complex conjugate zeros when $\Lambda <0$, (ii) Equation~(\ref{eq:relalf1}) has a multiple zero and all the zeros are real when $\Lambda = 0$, and (iii) Equation~(\ref{eq:relalf1}) has three distinct real zeros when $\Lambda >0$. This classification is very relevant because the complex zeros of Equation~(\ref{eq:relalf1}) result in damped oscillatory  solutions of Equation~(\ref{eq:relalf}) whereas the real zeros of Equation~(\ref{eq:relalf1})  correspond to  evanescent solutions of Equation~(\ref{eq:relalf}).

It is instructive to consider again the paradigmatic situation in which there are no collisions between the two fluids, so we set $\nuni=0$. The discriminant becomes $\Lambda= -4k_z^6 c_{\rm A}^6<0$, which means that Equation~(\ref{eq:relalf1}) has one real zero and two complex conjugate zeros. Indeed, when $\nuni=0$ the zeros of Equation~(\ref{eq:relalf1}) are
\begin{equation}
 s = \pm i k_z \ca, \qquad s=0, 
\end{equation}
which correspond to the following values of $\omega$,
\begin{equation}
 \omega = \pm k_z\ca, \qquad \omega=0. \label{eq:uncoupled}
\end{equation}
The two non-zero solutions correspond to the ideal Alfv\'en frequency, as expected.

We go back to the general case $\nuni \neq 0$. To determine the location where the nature of the solutions changes we set $\Lambda = 0$ and find the corresponding relation between the various parameters. For given $\nuni$ and $\chi$ we find two different values of $k_z$, denoted by $k_z^+$ and $k_z^-$, which satisfy $\Lambda = 0$, namely
\begin{equation}
k_z^\pm = \frac{\nu_{\rm ni}}{c_{\rm A}}\left[\frac{\chi^2+20\chi-8}{8\left( 1+\chi \right)^3} \pm \frac{\chi^{1/2} \left(\chi-8 \right)^{3/2}}{8\left( 1+\chi \right)^3}\right]^{-1/2}. \label{eq:rmasmenos}
\end{equation}
Since $k_z$ must be real, Equation~(\ref{eq:rmasmenos}) imposes a condition on the minimum value of $\chi$ which allows $\Lambda = 0$. This minimum value is $\chi = 8$ and the corresponding critical $k_z$ is $k_z^+ = k_z^- = 3\sqrt{3} \nuni/\ca$. When $\chi > 8$, Equation~(\ref{eq:rmasmenos}) gives $k_z^+ < k_z^-$. For $k_z$ outside the interval $(k_z^+,k_z^-)$ we have $\Lambda <0$ so that there are two propagating Alfv\'en waves and one evanescent solution.  For $k_z\in(k_z^+,k_z^-)$ we have $\Lambda >0$ and so all three zeros of Equation~(\ref{eq:relalf1}) are real, i.e., they correspond to purely imaginary solutions of Equation~(\ref{eq:relalf}). There is no propagation of Alfv\'en waves for $k_z\in(k_z^+,k_z^-)$. We call this interval the cut-off region. To the best of our knowledge, \citet{kulsrud1969} were the first to report on the existence of a cut-off region of wavenumbers for Alfv\'en waves in a partially ionized two-fluid plasma, when studying the propagation of cosmic rays. These cut-offs also appear in the works by, e.g., \citet{pudritz1990,kumar2003,mouschovias2011}. The cut-off wavenumbers were ignored by \citet{zaqarashvili2011a}, who  stated that there is always a solution of Equation~(\ref{eq:relalf}) with a non-zero real part.  \citet{zaqarashvili2011a} probably reached this wrong conclusion because they never took $\chi > 8$ in their computations. Here we clearly see that the three solutions of Equation~(\ref{eq:relalf}) are purely imaginary when $\chi>8$ and $k_z\in(k_z^+,k_z^-)$.

\citet{kamaya1998} discussed the physical reason for the existence of a range of cut-off wavenumbers in weakly ionized plasmas \citep[see also][]{mouschovias1987}. When $k_z > k_z^-$ magnetic tension drives ions to oscillate almost freely, since the friction force is not strong enough to transfer significant inertia to neutrals. In this case, disturbances in the magnetic field affect only the ionized fluid as happens for classic Alfv\'en waves in fully ionized plasmas. Conversely, when $k_z < k_z^+$ the ion-neutral friction is efficient enough for neutrals to be nearly frozen into the magnetic field. After a perturbation, neutrals are dragged by ions almost instantly and both species oscillate together as a single fluid. The intermediate situation occurs when $k_z\in(k_z^+,k_z^-)$. In this case, a disturbance in the magnetic field decays due to friction before the ion-neutral coupling has had time to transfer the restoring properties of magnetic tension to the neutral fluid. In other words, neutral-ion collisions are efficient enough to dissipate  perturbations in the magnetic field but, on the contrary, they are not efficient enough to transfer significant inertia to neutrals before the magnetic field perturbations have decayed. Hence, oscillations of the magnetic field are suppressed when $k_z\in(k_z^+,k_z^-)$. Additional insight on the physical behavior of the perturbations near the cut-off region  is given in Section~\ref{sec:compar} by analyzing the forces acting on the fluids.

To avoid confusion we must inform the reader that the existence of cut-off wavenumbers of Alfv\'en waves discussed above is a purely two-fluid effect.  These  cut-off wavenumbers are not  the cut-offs obtained in the single-fluid approximation \citep[see, e.g.,][]{balsara1996,forteza2007,soler2009PI,barcelo2011}. \citet{zaqarashvili2012} have shown that the cut-off wavenumbers found in the single-fluid approximation are a mathematical artifact, i.e., they are caused by the approximations made when proceeding from the multi-fluid equations to single-fluid equations and are not connected to any real physical process.  On the contrary, the cut-off wavenumbers found in the two-fluid case are physically and mathematically real and are caused by the two-fluid interaction between ions and neutrals \citep[see][]{pudritz1990,kamaya1998,mouschovias2011}.

\subsubsection{Approximate analytic solutions}

We look for approximate analytic solutions to Equation~(\ref{eq:relalf}) corresponding to standing modes. We assume that $k_z$ is outside the cut-off interval $(k_z^+,k_z^-)$, so that Equation~(\ref{eq:relalf}) has two complex solutions and one purely imaginary solution. This is the most interesting situation for the study of standing Alfv\'en waves since no oscillatory modes exist when $k_z$ is within the cut-off interval. First we look for an approximate expression for the two oscillatory solutions. To do so we write $\omega = \omega_{\rm R} + i \omega_{\rm I}$ and insert this expression in Equation~(\ref{eq:relalf}). We assume $|\omega_{\rm I}|\ll |\omega_{\rm R}|$ and neglect terms with $\omega_{\rm I}^2$ and higher powers. Hence, it is crucial for the validity of this approximation that $k_z$ is not within or close to the cut-off region where $\omega_{\rm R} =0$. After some algebraic manipulations we derive approximate expressions for $\omega_{\rm R}$ and $\omega_{\rm I}$. For simplicity we omit the intermediate steps and give the final expressions, namely
\begin{eqnarray}
 \omega_{\rm R} &\approx &  \pm k_z \ca \sqrt{\frac{k_z^2 c_{\rm A}^2+\left( 1 +\chi \right)\nuni^2}{ k_z^2 c_{\rm A}^2+\left( 1 +\chi \right)^2\nuni^2}}, \label{eq:wr} \\
\omega_{\rm I}&\approx &   -\frac{\chi \nuni}{2\left[  k_z^2 c_{\rm A}^2 +\left( 1+\chi \right)^2 \nuni^2\right]}  k_z^2 c_{\rm A}^2. \label{eq:wi}
\end{eqnarray}
On the other hand, the remaining purely imaginary, i.e., evanescent, solution is $\omega = i \epsilon$, with the approximation to $\epsilon$ given by
\begin{equation}
 \epsilon \approx - \nuni \frac{k_z^2 c_{\rm A}^2+\left( 1 +\chi \right)^2\nuni^2}{k_z^2 c_{\rm A}^2+\left( 1 +\chi \right)\nuni^2}. \label{eq:gamma}
\end{equation}
When $\nuni=0$, we find $\omega_{\rm R} =  \pm k_z c_{\rm A}$, $\omega_{\rm I} = 0$ and $\epsilon=0$, hence we recover the solutions in the uncoupled case (Equation~(\ref{eq:uncoupled})).

It is useful to investigate the behavior of the solutions in the various limits of $\nuni$. First we consider the limit $\nuni \ll k_z \ca$, i.e., the case of low collision frequency, which means that the coupling between fluids is weak. Equations~(\ref{eq:wr}) and (\ref{eq:gamma}) simplify to
\begin{eqnarray}
 \omega_{\rm R} &\approx &  \pm k_z \ca, \label{eq:wrlow} \\
\omega_{\rm I}&\approx &   -\frac{\chi \nuni}{2}, \label{eq:wilow} \\
\epsilon & \approx & -\nuni.
\end{eqnarray}
In this limit $\omega_{\rm R}$ coincides with its value in the ideal, uncoupled case and $\omega_{\rm I}$ is independent of $k_z$. Hence, the the damping of Alfv\'en waves does not depend on the wavenumber.

On the other hand, when $\nuni \gg k_z \ca$, i.e., the case of strong coupling between fluids,  we find
\begin{eqnarray}
 \omega_{\rm R} &\approx &  \pm \frac{k_z \ca}{\sqrt{1+\chi}}, \label{eq:wrhigh} \\
\omega_{\rm I}&\approx &   -\frac{\chi}{2\left( 1+\chi \right)^2 } \frac{k_z^2 c_{\rm A}^2 }{\nuni}, \label{eq:wihigh} \\
\epsilon & \approx & -(1+\chi)\nuni.
\end{eqnarray}
Now the expression of $\omega_{\rm R}$ involves the factor $\sqrt{1+\chi}$ in the denominator, so that the larger the amount of neutrals, the lower $\omega_{\rm R}$ compared to the value in the fully ionized case \citep[see also][]{kumar2003,soler2012}. Now $\omega_{\rm I}$ is proportional to $k_z^2$, meaning that the shorter the wavelength, the more efficient damping.

\subsubsection{Comparison with numerical results}
\label{sec:compar}

\begin{figure}
	\centering
	\includegraphics[width=.65\columnwidth]{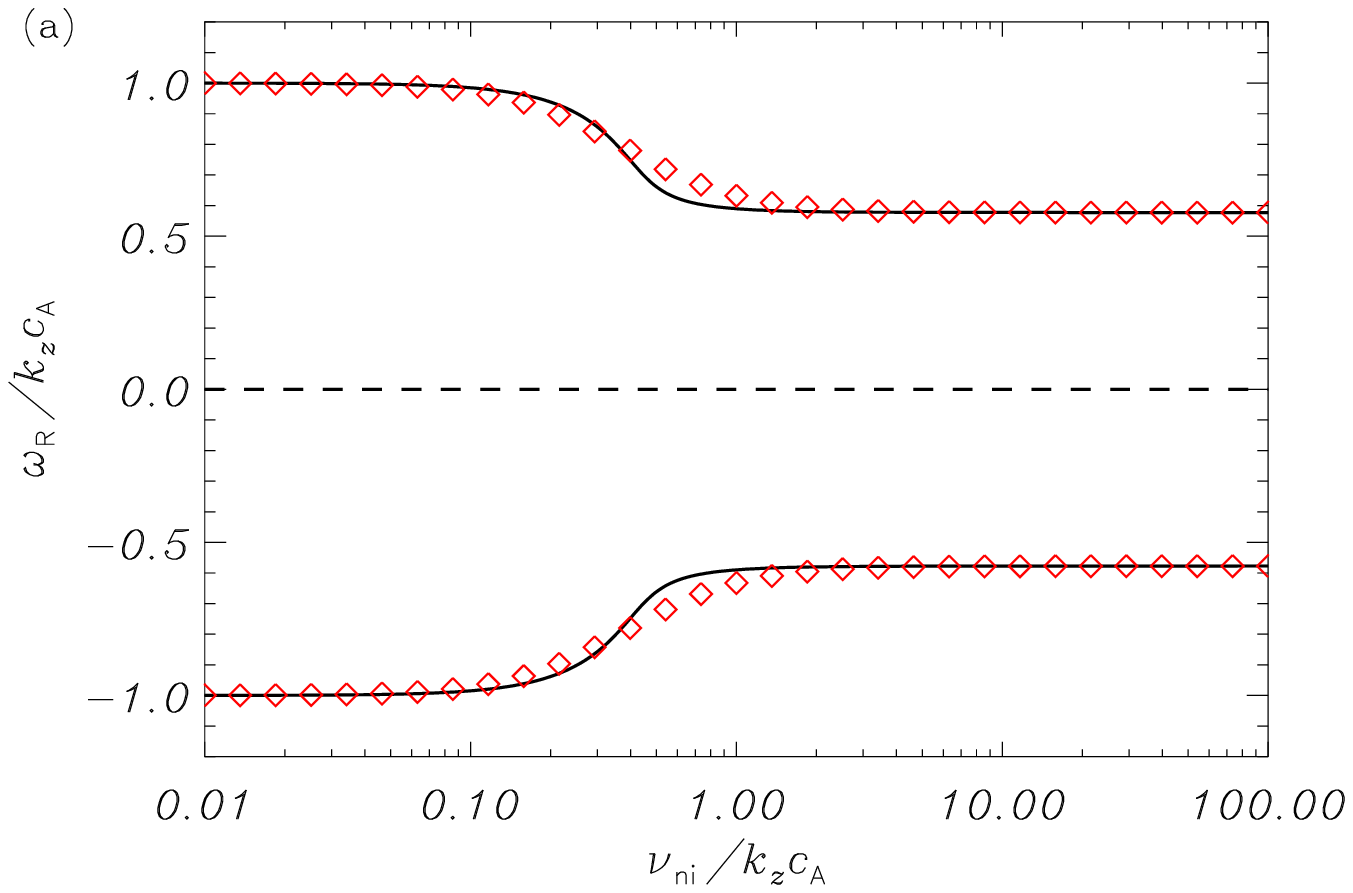}
	\includegraphics[width=.65\columnwidth]{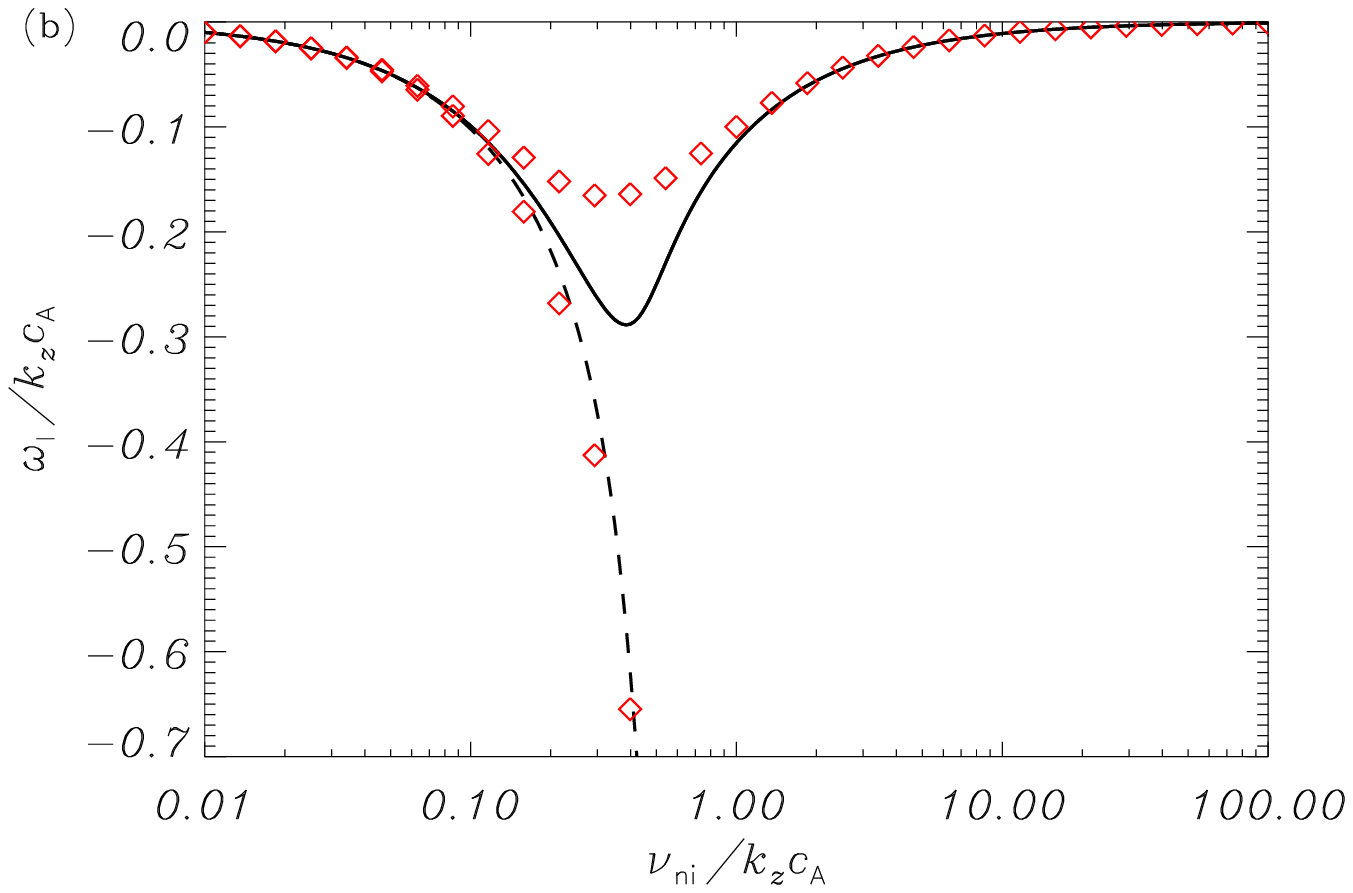}
	\caption{Results for standing waves. (a) $\omega_{\rm R}/k_z \ca$ and (b) $\omega_{\rm I}/k_z \ca$ as functions of $\nuni/k_z \ca$. We have used $\chi=2$. Solid and dashed lines correspond to the numerical results of the oscillatory and evanescent modes, respectively, while  the symbols correspond to the analytic expressions in the weak damping approximation (Equations~(\ref{eq:wr})--(\ref{eq:gamma})). }
	\label{fig:compara}
\end{figure}

Here we solve the full dispersion relation (Equation~(\ref{eq:relalf})) numerically and compare the numerical solutions with the previous approximations (Equations~(\ref{eq:wr})--(\ref{eq:gamma})). First we set  $\chi=2$ and vary the ratio $\nuni/k_z \ca$ between $10^{-2}$ and $10^2$. We compute $\omega_{\rm R}/k_z \ca$ and $\omega_{\rm I}/k_z \ca$ (see Figure~\ref{fig:compara}). The agreement between numerical and analytic results is very good. There is no cut-off region for this choice of parameters because we have taken $\chi<8$. Regarding the real part of the frequency, we obtain that the oscillatory modes have $\omega_{\rm R}/k_z \ca \approx \pm 1 $ when $\nuin / k_z \ca \ll 1$. When the ratio $\nuni / k_z \ca $ increases,  $\omega_{\rm R}/k_z \ca  $ decreases until de value $\omega_{\rm R}/ k_z \ca  \approx \pm 1/\sqrt{1+\chi}$ is reached. This behavior is consistent with the analytic Equation~(\ref{eq:wr}). The evanescent mode has $\omega_{\rm R} = 0$ regardless the value of $\nuni / k_z \ca $. The imaginary part of the frequency of the oscillatory modes tend to zero in both limits  $\nuni / k_z \ca  \ll 1$ and $\nuni / k_z \ca  \gg 1$, while the damping is most efficient when $\nuni/k_z \ca \sim 1$. This result is also consistent  with the analytic Equation~(\ref{eq:wi}), although the approximation underestimates the actual damping rate when $\nuni/k_z \ca \sim 1$. This discrepancy is a consequence of the weak damping approximation, which assumes $|\omega_{\rm I}|\ll |\omega_{\rm R}|$. However when $\nuni/k_z \ca \sim 1$, the numerical results show that $|\omega_{\rm I}|$ and $|\omega_{\rm R}|$ are of the same order, hence the damping is strong. The imaginary part of the frequency of the evanescent mode is very well approximated by Equation~(\ref{eq:gamma}).

\begin{figure}
	\centering
	\includegraphics[width=.65\columnwidth]{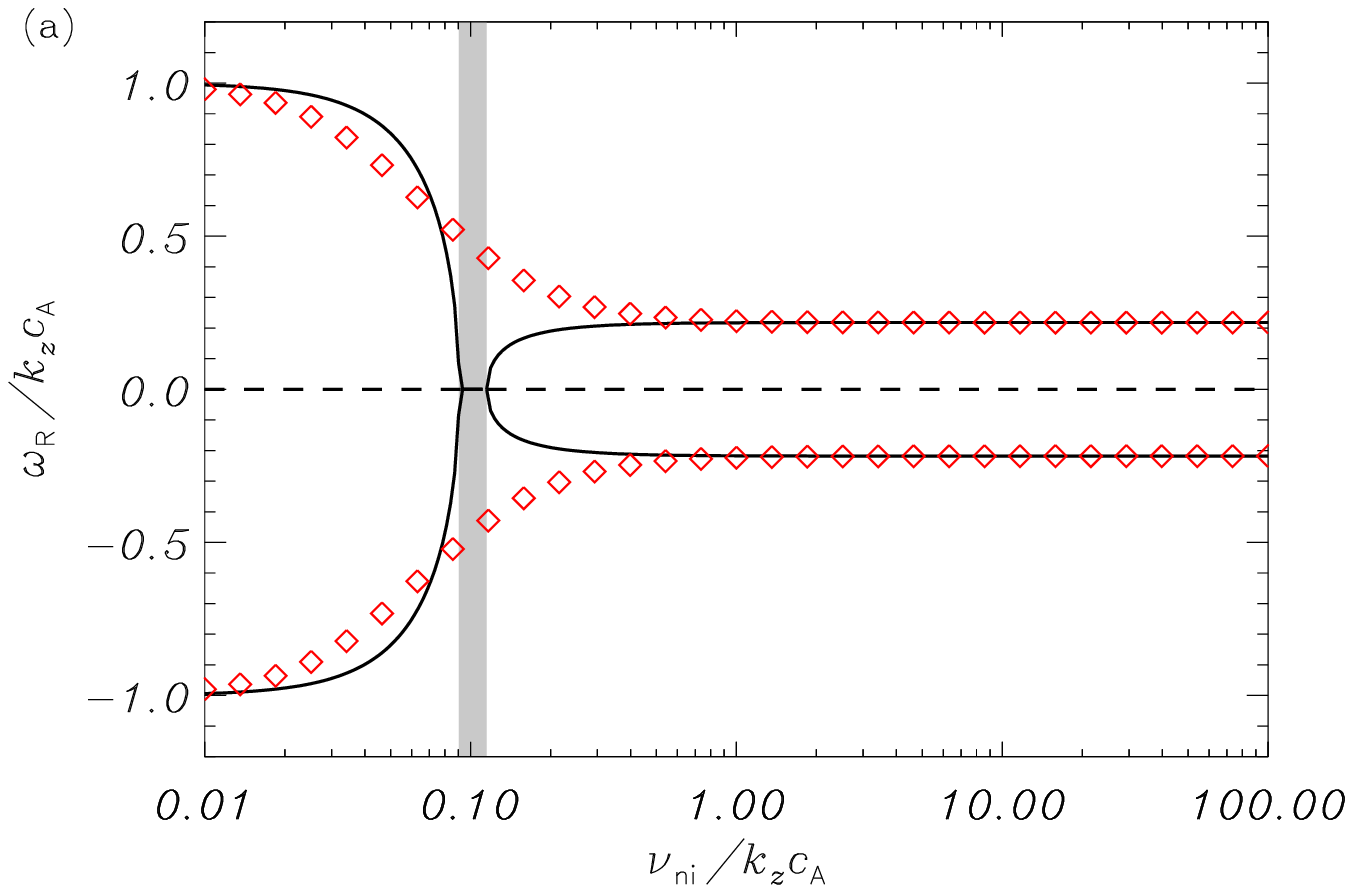}
	\includegraphics[width=.65\columnwidth]{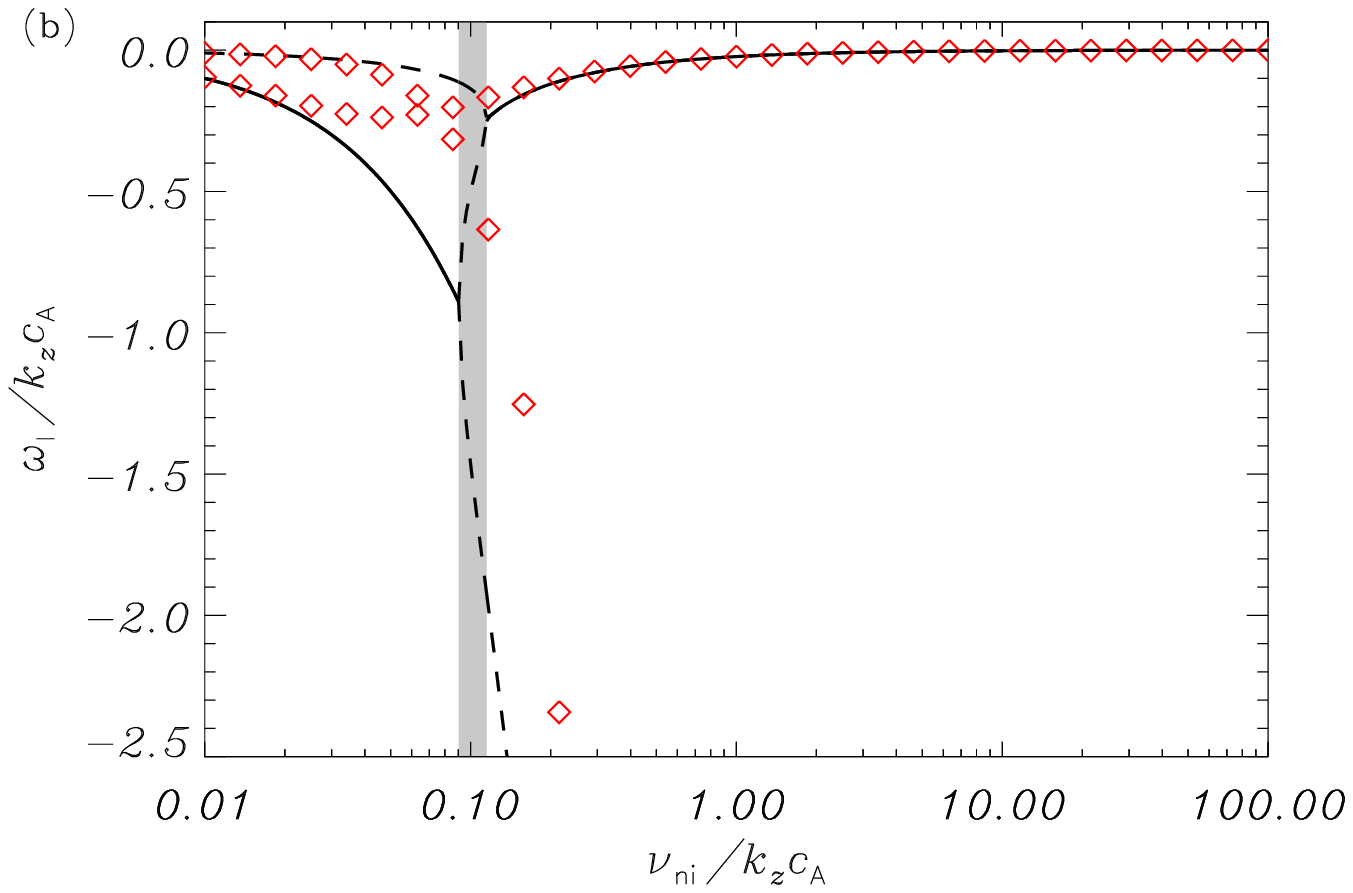}
	\caption{Same as  Figure~\ref{fig:compara} but with $\chi = 20$. The shaded zone denotes the cut-off region according to  Equation~(\ref{eq:rmasmenos}).}
	\label{fig:compara2}
\end{figure}

Next we increase ionization ratio to $\chi = 20$ and compute the same results as before (Figure~\ref{fig:compara2}). Now there is a cut-off region because $\chi > 8$. The cut-off region is correctly described by Equation~(\ref{eq:rmasmenos}). At the cut-off the oscillatory and evanescent modes interact and they become three purely imaginary solutions. Propagation is forbidden in this interval.  As expected, the agreement between numerical and analytic results is not good near the cut-off region, but both results are in reasonably agreement far from the cut-off interval.

\begin{figure}
	\centering
	\includegraphics[width=.65\columnwidth]{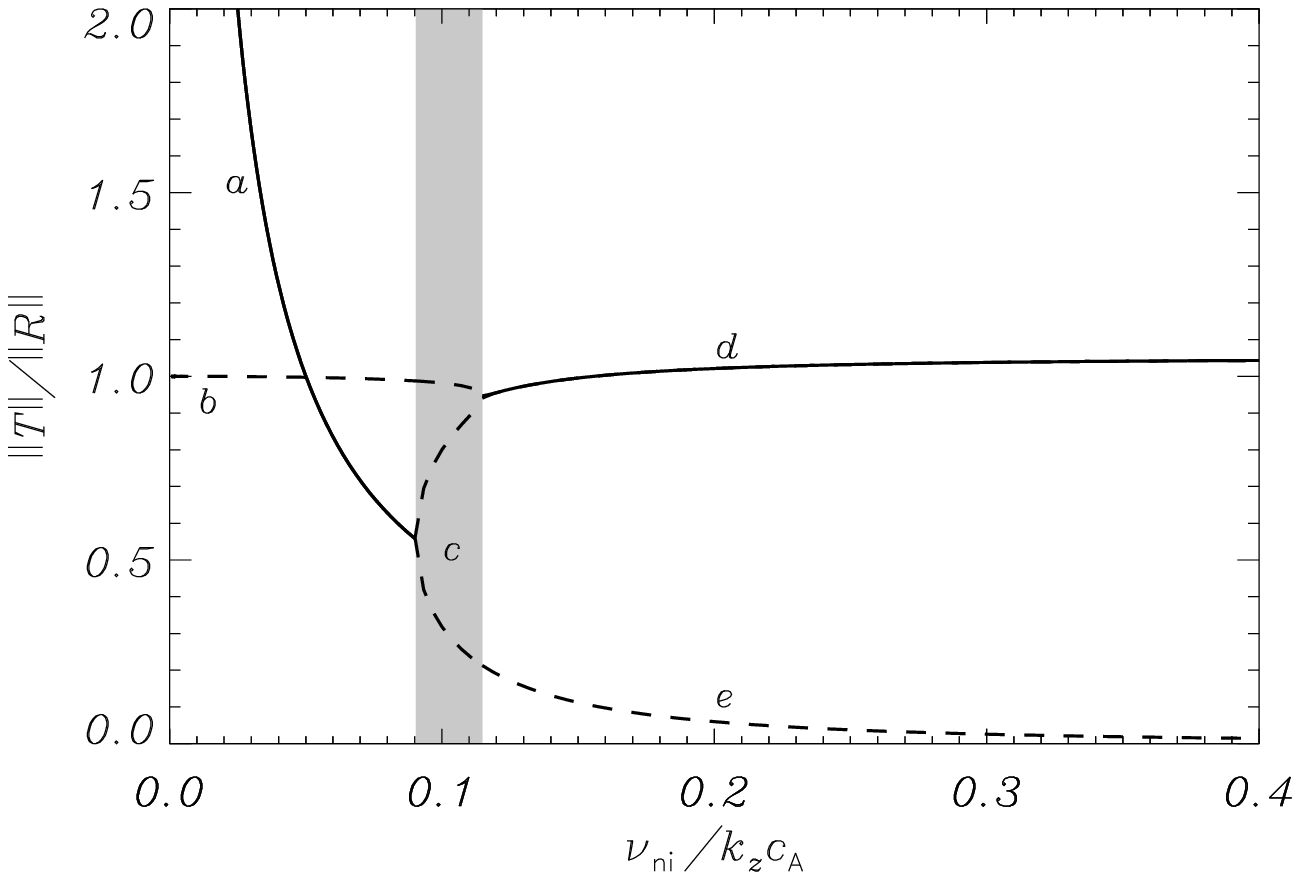}
	\caption{Ratio $||{\bf T}||/||{\bf R}||$ versus $\nuni/k_z \ca$ for the solutions displayed in Figure~\ref{fig:compara2} near the cut-off region (shaded zone). Solid and dashed lines correspond to oscillatory and evanescent solutions in time, respectively.}
	\label{fig:modulus}
\end{figure}

To explore the physical behavior of the solutions near the cut-off region, we rewrite the momentum equations of ions (Equation~(\ref{eq:momlinion})) and neutrals (Equation~(\ref{eq:momlinneu}))  in the following forms,
\begin{eqnarray}
\rhoi \frac{\pd {\bf v}_{\rm i}}{\partial t} &=& {\bf T} - {\bf R}, \label{eq:momion2} \\
\rhon \frac{\pd {\bf v}_{\rm n}}{\partial t} &=& {\bf R}, \label{eq:momneu2}
\end{eqnarray}
where $\bf T$ and $\bf R$ are the magnetic tension force and the friction force, respectively, given by
\begin{eqnarray}
{\bf T} &=& - i \rhoi \frac{k_z^2 \ca^2}{\omega} {\bf v}_{\rm i}, \\
{\bf R} &=& \rhon \frac{\nuni\omega}{\omega + i\nuni} {\bf v}_{\rm i}.
\end{eqnarray}
In Equations~(\ref{eq:momion2}) and (\ref{eq:momneu2}) we have not included magnetic pressure and gas pressure forces because they do not affect Alfv\'en waves. Now we use the numerically obtained solutions for $\chi = 20$ (Figure~\ref{fig:compara2}) to compute the moduli of $\bf T$ and $\bf R$, namely $||{\bf T}||$ and $||{\bf R}||$, as functions of $\nuni/k_z \ca$. Figure~\ref{fig:modulus} displays the ratio $||{\bf T}||/||{\bf R}||$ versus $\nuni/k_z \ca$ near the cut-off region. We have selected some locations in Figure~\ref{fig:modulus}, denoted by letters from {\em a} to {\em e}, to support the following discussion on the importance of the two forces.

We start by analyzing the solutions on the left-hand side to the cut-off region. There are an oscillatory solution, {\em a}, and an evanescent solution, {\em b}. We find that $||{\bf T}|| \gg ||{\bf R}||$ for the oscillatory solution {\em a}, so that there is a net restoring force for ions in Equation~(\ref{eq:momion2}). Magnetic tension is the dominant force and drives ions to oscillate almost freely, whereas neutrals are only slightly perturbed by the weak friction force in Equation~(\ref{eq:momneu2}).  In the case of the evanescent solution {\em b} we obtain that $||{\bf T}|| \approx ||{\bf R}||$. This means that there is no net force acting on ions. The evanescent solution {\em b} only produces perturbations in the neutral fluid.

We turn to  location {\em c} in Figure~\ref{fig:modulus}, i.e., within the cut-off interval. Here all the solutions are evanescent. The ratio $||{\bf T}||/||{\bf R}||$ of the solution that was previously oscillatory decreases  and becomes $||{\bf T}||/||{\bf R}|| < 1$ before reaching the cut-off region. Now friction is the dominant force. Friction acts very efficiently in dissipating  perturbations in the plasma before ions (and neutrals indirectly) have had time to feel the restoring force of magnetic tension. As a consequence, oscillatory modes are suppressed.

Finally, we analyze the forces acting on the solutions on the right-hand side to the cut-off region. Again, there are an oscillatory solution, {\em d}, and an evanescent solution, {\em e}. As happened for the oscillatory solution {\em a}, we find that $||{\bf T}|| > ||{\bf R}||$ for the oscillatory solution {\em d}, although in this case the friction force is not negligible. Magnetic tension provides now the necessary restoring force for the oscillations of ions, while the friction force  is responsible for dragging neutrals when ions move. Hence, both species tend to oscillate together. Solution {\em d} represents a collective oscillation of the whole plasma. On the contrary, magnetic tension is negligible for the evanescent solution, {\em e}. This mode is governed by the friction force alone and simply causes the decay of perturbations.

\subsection{Propagating waves}

We move to the study of propagating waves. In the ideal fully ionized case the study of propagating waves is equivalent to that of standing waves. Here we shall see that ion-neutral collisions break this equivalence and propagating waves are worth being studied separately. For propagating waves we assume a real $\omega$ and solve the dispersion relation (Equation~(\ref{eq:relalf})) to find the complex wavenumber, $k_z = k_{z,\rm R}+i k_{z,\rm I}$, with $k_{z,\rm R}$ and $k_{z,\rm I}$ the real and imaginary parts of $k_z$, respectively. Then the amplitude of perturbations is multiplied by the factor $\exp(-k_{z,\rm I} z)$, so that the perturbations are spatially damped. For real and positive $\omega$, $k_{z,\rm R} > 0$ corresponds to waves propagating towards the positive $z$-direction. Conversely, $k_{z,\rm R} < 0$ corresponds to waves propagating towards the negative $z$-direction. In both situations, the sign of $k_{z,\rm I}$ is the same as that of $k_{z,\rm R}$.

Equation~(\ref{eq:relalf}) is a quadratic equation in $k_z$.  The solution to Equation~(\ref{eq:relalf})  is
\begin{equation}
k_z^2 = \frac{\omega^2}{\ca^2} \frac{\omega + i (1+\chi)\nuni}{\omega+i\nuni}. \label{eq:kz2}
\end{equation}
When $\nuni=0$ we recover the ideal result $k_z^2 = \omega^2/\ca^2$. For $\nuni\neq 0$ we write $k_z = k_{z,\rm R}+i k_{z,\rm I}$ and insert this expression in Equation~(\ref{eq:kz2}). Then, it is possible to obtain the exact expression of $k_{z,\rm R}^2$, namely
\begin{eqnarray}
k_{z,\rm R}^2 &=& \frac{1}{2} \frac{\omega^2}{\ca^2} \frac{\omega^2 + (1+\chi)\nuni^2}{\omega^2 + \nuni^2} \nonumber \\
&\times & \left[ 1 + \left( 1 + \frac{\chi^2\nuni^2\omega^2}{(\omega^2 + (1+\chi)\nuni^2)^2} \right)^{1/2} \right],\label{eq:kzr2}
\end{eqnarray}
while the exact expression of $k_{z,\rm I}^2$ is
\begin{equation}
k_{z,\rm I}^2 = k_{z,\rm R}^2 - \frac{\omega^2}{\ca^2} \frac{\omega^2 + (1+\chi)\nuni^2}{\omega^2 + \nuni^2}.\label{eq:kzi2}
\end{equation}

Contrary to the case of standing waves there is no cut-off region for propagating Alfv\'en waves. It is always found that $k_{z,\rm R} \neq 0$ regardless the value of $\omega$. This apparent contradiction between the standing and propagating cases can be understood as follows. An oscillatory standing wave can be interpreted as the superposition of two  propagating waves with the same frequency running in opposite directions. However such a representation does not work for a  perturbation fixed in space and evanescent in time. The presence of static evanescent perturbations is a peculiar result only obtained when $k_z$ is real and $\omega$ is purely imaginary, a situation that cannot be described with propagating waves. For this reason evanescent solutions are absent from the present study of propagating waves. 

\subsubsection{Approximate expressions and comparison with numerical results}

Going back to Equations~(\ref{eq:kzr2}) and (\ref{eq:kzi2}), we realize that it is possible to obtain simpler expressions of $k_{z,\rm R}^2$ and $k_{z,\rm I}^2$ by taking advantage of the fact that the second term within the square root of Equation~(\ref{eq:kzr2}) is always much smaller than unity. We can approximate Equations~(\ref{eq:kzr2}) and (\ref{eq:kzi2}) as
\begin{eqnarray}
k_{z,\rm R}^2 &\approx& \frac{\omega^2}{\ca^2} \frac{\omega^2 + (1+\chi)\nuni^2}{\omega^2 + \nuni^2} \nonumber \\ &+& \frac{1}{4} \frac{\omega^2}{\ca^2} \frac{\chi^2\nuni^2\omega^2}{\left(\omega^2 + \nuni^2\right)\left(\omega^2 + (1+\chi)\nuni^2\right)},  \label{eq:appkr} \\
k_{z,\rm I}^2 &\approx& \frac{1}{4} \frac{\omega^2}{\ca^2} \frac{\chi^2\nuni^2\omega^2}{\left(\omega^2 + \nuni^2\right)\left(\omega^2 + (1+\chi)\nuni^2\right)}.\label{eq:appki}
\end{eqnarray}
As for standing waves we evaluate the analytic solutions (Equations~(\ref{eq:appkr}) and (\ref{eq:appki})) in the various limits of $\nuni$. When $\nuni \ll \omega$, Equations~(\ref{eq:appkr}) and (\ref{eq:appki}) simplify to
\begin{eqnarray}
 k_{z,\rm R} &\approx & \pm \sqrt{ \frac{\omega^2}{\ca^2} + \frac{\chi^2\nuni^2}{4\ca^2}} \approx \pm \frac{\omega}{\ca}, \label{eq:krlow} \\
k_{z,\rm I} &\approx &  \pm \frac{\chi\nuni}{2\ca}. \label{eq:kilow} 
\end{eqnarray}
Hence we find that $ k_{z,\rm R}$ takes the same value as in the fully ionized case, and $k_{z,\rm I}$ is independent of $\omega$. Conversely when $\nuni \gg \omega$  we find
\begin{eqnarray}
 k_{z,\rm R} &\approx & \pm \sqrt{ \frac{\omega^2}{\ca^2}(1+\chi) + \frac{1}{4}\frac{\omega^2}{\ca^2}\frac{\chi^2\omega^2}{(1+\chi)^2\nuin^2} } \nonumber \\ &\approx& \pm \frac{\omega}{\ca}\sqrt{1+\chi}, \label{eq:krhigh} \\
k_{z,\rm I} &\approx & \pm \frac{\chi\omega^2}{2(1+\chi)\ca\nuin}. \label{eq:kihigh}
\end{eqnarray}
We find the presence of the factor $\sqrt{1+\chi}$ in the approximate expression of $k_{z,\rm R}$, while now $k_{z,\rm I}$ is proportional to $\omega^2$. Hence, high-frequency waves are more efficiently damped than low-frequency waves.

\begin{figure}
	\centering
	\includegraphics[width=.65\columnwidth]{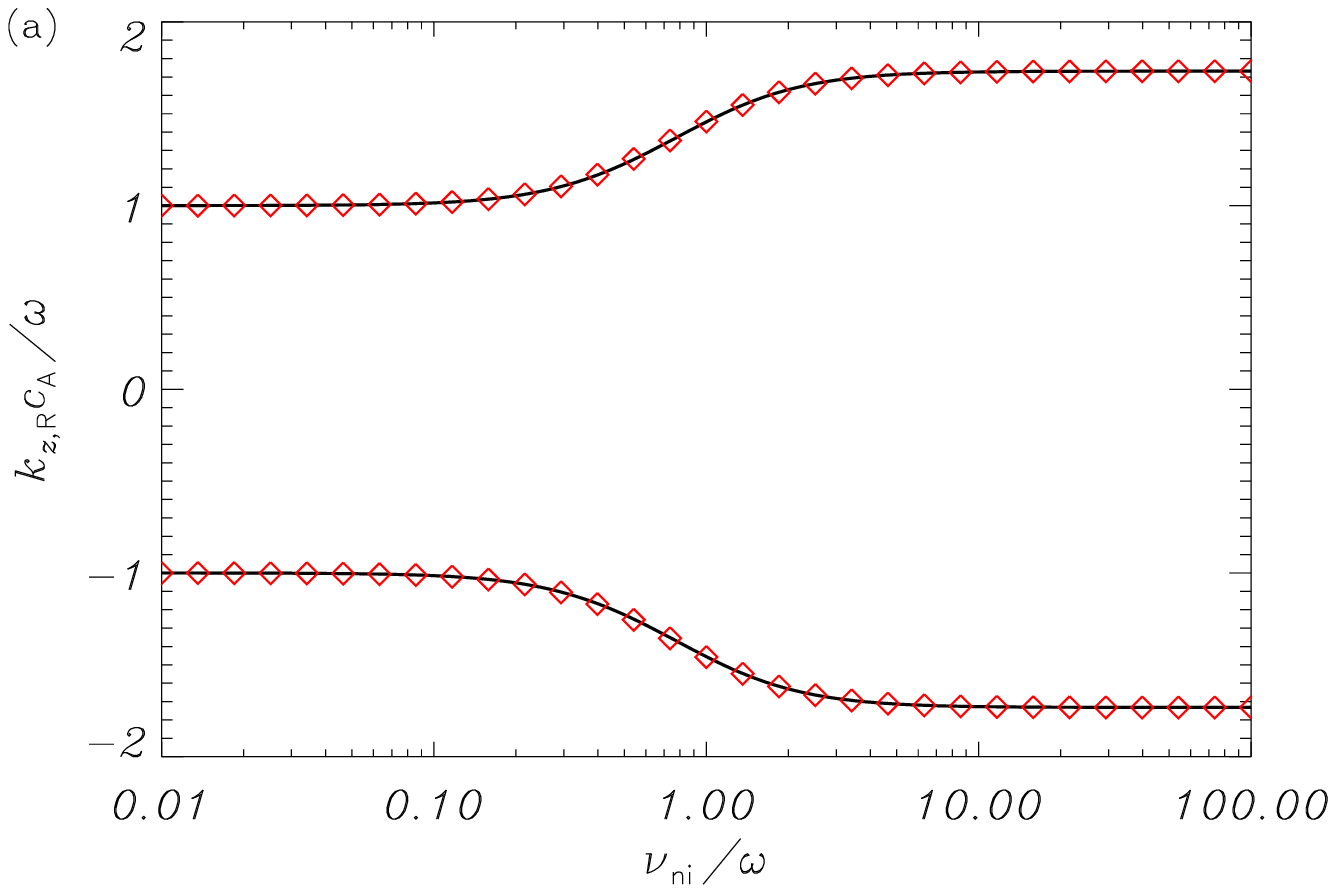}
	\includegraphics[width=.65\columnwidth]{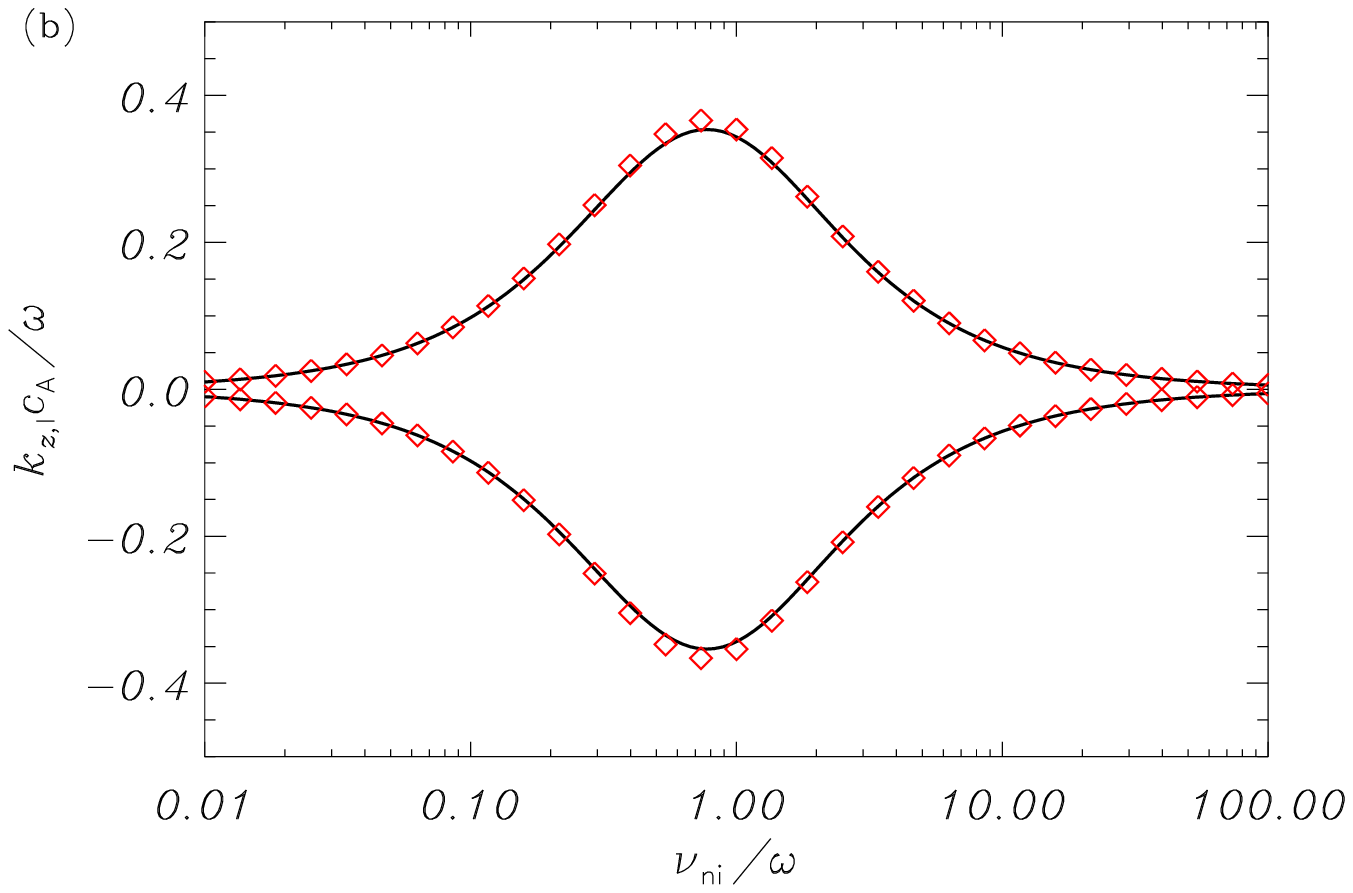}	
	\caption{Results for propagating waves. (a) $k_{z,\rm R}\ca/\omega$ and (b) $k_{z,\rm I}\ca/\omega$ as functions of $\nuni/\omega$.  Solid lines correspond to the numerical results while symbols correspond to the analytic approximations (Equations~(\ref{eq:appkr}) and (\ref{eq:appki})). We have used $\chi=2$. }
	\label{fig:compara3}
\end{figure}

Now we solve the dispersion relation (Equation~(\ref{eq:relalf})) numerically and compare the numerical solutions with the analytic approximations of $k_{z,\rm R}$ and $k_{z,\rm I}$ (Equations~(\ref{eq:appkr}) and (\ref{eq:appki})).  These results and shown in Figure~\ref{fig:compara3} for a particular set of parameters given in the caption of the Figure. As in the standing case, the agreement between numerical and analytic results is very good. We have replied these computations using various values of $\chi$ (not shown here for simplicity) and the analytic results are always in accordance with the numerical ones.

\subsubsection{Phase velocity and group velocity}

Here we compute the phase and group velocities of the propagating Alfv\'en waves. The phase velocity, ${\bf v}_{\rm ph}$, is defined as
\begin{equation}
{\bf v}_{\rm ph} = \frac{\omega}{k} \hat{e}_k,
\end{equation}
with $\hat{e}_k$ denoting the unit vector in the direction of the wave vector. From the dispersion relation (Equation~(\ref{eq:relalf})) we get
\begin{equation}
\frac{\omega}{k} = \ca\cos\theta \sqrt{\frac{\omega+i\nuni}{\omega+i(1+\chi)\nuni}}. \label{eq:phasevel}
\end{equation}
The phase velocity is a complex quantity. Its real part is related to the propagation speed of the wave while its imaginary part is related to the wave attenuation. Both real and imaginary parts of ${\bf v}_{\rm ph}$ have the same dependence on $\theta$, meaning that the damping rate is independent of the direction of propagation. Ion-neutral collisions do not affect the dependence of ${\bf v}_{\rm ph}$  on $\theta$. From Equation~(\ref{eq:phasevel}) we can define the effective Alfv\'en velocity, $\tilde{c}_{\rm A}$, as
\begin{equation}
\tilde{c}_{\rm A} = \ca \sqrt{\frac{\omega+i\nuni}{\omega+i(1+\chi)\nuni}}. \label{eq:effva}
\end{equation}

The group velocity, ${\bf v}_{\rm gr}$, is the propagation velocity of a wave packet and, therefore, of the wave energy. It is defined as
\begin{equation}
{\bf v}_{\rm gr} = \nabla_{\bf k} \omega = \frac{\partial \omega}{\partial k_x} \hat{e}_x + \frac{\partial \omega}{\partial k_y} \hat{e}_y  + \frac{\partial \omega}{\partial k_z} \hat{e}_z.
\end{equation}
Using the dispersion relation (Equation~(\ref{eq:relalf})) the group velocity can be more easily computed as
\begin{equation}
{\bf v}_{\rm gr} = - \frac{1}{\partial \mathcal{D} / \partial \omega} \left( \frac{\partial \mathcal{D}}{\partial k_x} \hat{e}_x + \frac{\partial \mathcal{D}}{\partial k_y} \hat{e}_y  + \frac{\partial \mathcal{D}}{\partial k_z} \hat{e}_z \right),
\end{equation}
that gives
\begin{eqnarray}
{\bf v}_{\rm gr} &=& \frac{2 k_z \ca^2 (\omega + i\nuni)}{3\omega^2 - k_z^2\ca^2 + i2(1+\chi)\nuni\omega} \hat{e}_z \nonumber \\
&=& 2 \ca \frac{\left(\omega+i\nuni  \right)^{3/2}\left( \omega + i(1+\chi)\nuni \right)^{1/2}}{2\omega^2-2(1+\chi)\nuni^2+i(1+\chi)\nuni\omega} \hat{e}_z.
\end{eqnarray}
As it happens for the phase velocity, the group velocity is also complex. Since the damping of propagating Alfv\'en waves depends on $\omega$, the components of the wave packet with higher $\omega$ are more damped than the components with low $\omega$. Hence the concept of the group velocity as the velocity at which the whole wave packet propagates becomes obsolete when there is dissipation.  However, following \citet{muschietti1993} it is still possible to define an effective group velocity as the velocity at which the center of the wave packet propagates. This effective group velocity is a time-dependent combination of the real and imaginary parts of  ${\bf v}_{\rm gr}$ and depends on the particular form of the wave packet \citep[see extensive details in][]{muschietti1993}.

In the limits of low and high collision frequency the damping is weak, so that the imaginary part of ${\bf v}_{\rm gr}$ can be neglected compared to the real part. When $\nuni \ll \omega$, we find that ${\bf v}_{\rm gr} \approx \ca\hat{e}_z$ as in the ideal case, while when $\nuni \gg \omega$, the group velocity is ${\bf v}_{\rm gr} \approx \ca/ \sqrt{1+\chi}\hat{e}_z$. Note again the presence of the factor $\sqrt{1+\chi}$.

\section{Initial-value problem}

\label{sec:initial}

In Section~\ref{sec:normal} we have followed a normal mode approach and have assumed that the temporal dependence of the perturbations is of the form $\exp \left( -i\omega t \right)$. Here we go beyond the normal mode analysis and look for general time-dependent solutions to Equations~(\ref{eq:vorti}) and (\ref{eq:vortn}). To this end we follow two different approaches. First we solve the time dependent problem analytically by means of the Laplace transform. Later we use the numerical code MolMHD \citep{bona2009} to numerically evolve the full set of basic Equations~(\ref{eq:momlinion})--(\ref{eq:presslin}). Finally, these two independent results are compared. In this section we restrict ourselves to standing waves and  assume a real and positive $k_z$. We refer the reader to \citet{roberge2007} for the problem of driven propagating waves in a two-fluid plasma.

\subsection{Analytic solution}

Here we look for general time-dependent solutions to the coupled Equations~(\ref{eq:vorti}) and (\ref{eq:vortn}) using the Laplace transform. We compute the Laplace transform in time of the vorticity perturbations $\gammai$ and $\gamman$ as
\begin{equation}
 \tilde{\Gamma}_{\beta} ( s ) = \mathcal{L}\left[ \Gamma_\beta \right] = \int_0^\infty \Gamma_\beta (t) \exp\left( -s t \right) \der t,
\end{equation}
with $s$ is the transformed variable and $\beta$ denotes either `i' or `n'. The Laplace transforms of the temporal derivatives of vorticity are given by
\begin{eqnarray}
 \mathcal{L}\left[ \frac{\partial \Gamma_\beta}{\partial t} \right] &=& s \tilde{\Gamma}_{\beta} ( s )  - \Gamma_{0,\beta}, \\
 \mathcal{L}\left[ \frac{\partial^2 \Gamma_\beta}{\partial t^2} \right] &=& s^2 \tilde{\Gamma}_{\beta} ( s )  - s \Gamma_{0,\beta} - \Gamma'_{0,\beta},
\end{eqnarray}
where $\Gamma_{0,\beta}$ and $\Gamma'_{0,\beta}$ denote the value of $\Gamma_\beta$ and its temporal derivative at $t=0$, respectively. From  Equations~(\ref{eq:vorti}) and (\ref{eq:vortn}) we deduce that the temporal derivatives of the vorticity perturbations at $t=0$ satisfy
\begin{eqnarray}
\Gamma'_{0,\rm i} &=& -\chi \nuni \left( \Gamma_{0,\rm i}  - \Gamma_{0,\rm n}  \right), \\
\Gamma'_{0,\rm n}  &=& \nuni \left( \Gamma_{0,\rm i} - \Gamma_{0,\rm n}  \right).
\end{eqnarray}
At the present stage, we leave $\Gamma_{0, \rm i}$ and $\Gamma_{0,\rm n} $ unspecified.

We apply the Laplace transform to Equations~(\ref{eq:vorti}) and (\ref{eq:vortn}). From the transformed Equation~(\ref{eq:vortn}) we can express $\tgamman$ in terms of $\tgammai$ as
\begin{equation}
\tgamman(s) = \frac{1}{s + \nuni} \left( \nuni \tgammai(s) + \Gamma_{0, \rm n} \right). \label{eq:tgamman}
\end{equation}
We insert this expression in the transformed Equation~(\ref{eq:vorti}) and  obtain the expression for $\tgammai$ as
\begin{equation}
  \tgammai(s) = \frac{\mathcal{B}(s)}{\mathcal{D}^*(s)}, \label{eq:tgammai}
\end{equation}
where the functions $\mathcal{B}(s)$ and $\mathcal{D}^*(s)$ are defined as 
\begin{eqnarray}
\mathcal{B}(s) &=& s \left[\left( s + \nuni \right)\Gamma_{0, \rm i} + \chi \nuni \Gamma_{0, \rm n}   \right], \\
 \mathcal{D}^*(s) &=& s^3 + \left( 1+\chi \right) \nuni s^2 + k_z^2\ca^2 s + \nuni k_z^2\ca^2, \label{eq:disp2}
\end{eqnarray}
The function $\mathcal{B}(s)$ depends on the initial conditions and so it contains information about how the waves are excited. Conversely, the function $\mathcal{D}^*(s)$ is independent on the initial conditions. The function $\mathcal{D}^*(s)$ plays a very important role because it corresponds to the dispersion function. It tells us how the plasma behaves after the excitation has taken place. We note that the expression of $\mathcal{D}^*(s)$ (Equation~(\ref{eq:disp2})) coincides with the left-hand side of Equation~(\ref{eq:relalf1}), which was obtained from the normal mode dispersion relation (Equation~(\ref{eq:relalf})) after performing the change of variable $\omega = -i s$. Hence the normal modes are consistently recovered from the present Laplace transform approach by setting $\mathcal{D}^*(s) = 0$. 

To compute the  vorticity perturbation, $\gammai$, in the actual temporal domain we  perform the inverse Laplace transform of Equation~(\ref{eq:tgammai}), namely
\begin{equation}
 \gammai(t) = \mathcal{L}^{-1} \left[  \tgammai(s)  \right] = \mathcal{L}^{-1} \left[  \frac{\mathcal{B}(s)}{\mathcal{D}^*(s)} \right]. \label{eq:tgammaiinv}
\end{equation} 
The inverse Laplace transform of Equation~(\ref{eq:tgammaiinv}) can be evaluated by using the technique of partial fraction decomposition \citep[see, e.g.,][]{dyke1999} if the roots of the equation $\mathcal{D}^*(s) = 0$ are known. The zeros of $\mathcal{D}^*(s)$ are precisely the normal modes studied in Section~\ref{sec:normal}, so that we can take advantage of the approximate expressions found in  Section~\ref{sec:normal}. Once $\gammai(t)$ is known, the corresponding expression of $\gamman(t)$ can be obtained using Equation~(\ref{eq:tgamman}) and applying the convolution theorem, namely
\begin{equation}
 \gamman(t) = \Gamma_{0, \rm n}  e^{-\nuni t} + \nuni \int_0^t \gammai(\tau) \exp\left[-\nuni\left( t-\tau \right)  \right] \der \tau.\label{eq:tgammaninv}
\end{equation}
From Equations~(\ref{eq:tgammaiinv}) and (\ref{eq:tgammaninv}) we obtain $\gammai(t)$ and $\gamman(t)$, namely
\begin{eqnarray}
\gammai(t) &=& A_1 \exp\left( \epsilon t \right) \nonumber \\ 
&+& \left[ A_2 \cos\left(\omega_{\rm R} t\right) + A_3 \sin\left(\omega_{\rm R} t\right) \right]\exp\left( \omega_{\rm I} t \right),\label{eq:finalgi}\\
 \gamman(t) &=& C_1 \exp\left( \epsilon t \right) + C_2 \exp\left( -\nuni t \right) \nonumber \\
  & +& \left[ C_3 \cos\left(\omega_{\rm R} t\right) + C_4 \sin\left(\omega_{\rm R} t\right) \right]\exp\left( \omega_{\rm I} t \right),\label{eq:finalgn}
\end{eqnarray}
with the approximate expressions of $\omega_{\rm R}$, $\omega_{\rm I}$, and $\epsilon$ given in Equations~(\ref{eq:wr})--(\ref{eq:gamma}) and the coefficients $A_1$--$A_3$ and $C_1$--$C_4$  given in the Appendix. In the expression of $\omega_{\rm R}$ (Equation~(\ref{eq:wr})) the $+$ sign has to be taken. We must recall that the expressions of $\omega_{\rm R}$, $\omega_{\rm I}$, and $\epsilon$ given in Equations~(\ref{eq:wr})--(\ref{eq:gamma}) are computed in the weak damping approximation.

In this analysis we have worked with the vorticity perturbations. However we can also write the solution to the initial-value problem using velocity perturbations. To do so we move, for convenience, to a reference frame in which $k_y = 0$. Thus, the motions of Alfv\'en waves are polarized in the $y$-direction and from Equations~(\ref{eq:defgi}) and (\ref{eq:defgn}) we get that the vorticity perturbations are proportional to the $y$-component of velocities, namely $v_{{\rm i},y}$ and $v_{{\rm n},y}$. Therefore the expressions for the temporal evolution of $v_{{\rm i},y}$ and $v_{{\rm n},y}$ are the same as those given  in Equations~(\ref{eq:finalgi}) and (\ref{eq:finalgn}) for $\gammai$ and $\gamman$, respectively. We just need to replace $\Gamma_{0, \rm i}$ by  $ v_{{\rm i},0}$ and $\Gamma_{0, \rm n}$ by $v_{{\rm n},0}$, where $ v_{{\rm i},0}$ and  $v_{{\rm n},0}$ denote the values of the ion and neutral velocities at $t=0$, respectively.

\subsubsection{Solution in the absence of magnetic field}

We start the study of some interesting cases by considering the situation in which there is no magnetic field. We set $B=0$, and therefore $\ca=0$. This situation was studied by \citet{vranjes2008}. Equations~(\ref{eq:finalgi}) and (\ref{eq:finalgn}) become
\begin{eqnarray}
\gammai(t) &=& \frac{\Gamma_{0, \rm i} + \chi \Gamma_{0, \rm n}}{1+\chi} \nonumber \\ &+& \frac{\chi \left( \Gamma_{0, \rm i} - \Gamma_{0, \rm n} \right) }{1+\chi}\exp \left[ - \left( 1 + \chi \right) \nuni t \right], \label{eq:solb0i} \\
\gamman(t) &=& \frac{\Gamma_{0, \rm i} + \chi \Gamma_{0, \rm n}}{1+\chi} \nonumber \\ &-& \frac{\left( \Gamma_{0, \rm i} - \Gamma_{0, \rm n} \right) }{1+\chi}\exp \left[ - \left( 1 + \chi \right) \nuni t \right]. \label{eq:solb0n}
\end{eqnarray}
These solutions agree with those found by \citet{vranjes2008}. After an initial phase, i.e., when $\left( 1 + \chi \right) \nuni t \gg 1$, both $\gammai$ and $\gamman$ relax to the same constant value, which corresponds to a weighted average of the initial conditions, namely
\begin{equation}
\gammai(t) \approx \gamman(t) \approx \frac{\Gamma_{0, \rm i} + \chi \Gamma_{0, \rm n}}{1+\chi} = \frac{\rhoi\Gamma_{0, \rm i} + \rhon \Gamma_{0, \rm n}}{\rhoi+\rhon} . \label{eq:amplitude}
\end{equation} 
In the absence of magnetic field no oscillatory solutions are found. Vorticity perturbations remain constant after the relaxation phase.

\subsubsection{Low collision frequency}

Here we incorporate again the magnetic field but we consider the case of low collision frequency compared to the oscillation frequency, i.e., $\nuni \ll \omega_{\rm R}$.  Equations~(\ref{eq:finalgi}) and (\ref{eq:finalgn}) simplify to
\begin{eqnarray}
\gammai(t) & \approx & \Gamma_{0, \rm i} \cos \left( k_z \ca t \right) \exp \left( - \frac{\chi \nuni}{2} t \right), \label{eq:gilow} \\
 \gamman(t) & \approx & \Gamma_{0, \rm n} \exp \left( -  \nuni t \right). \label{eq:gnlow}
\end{eqnarray}
When $\nuni \ll \omega_{\rm R}$ the vorticity perturbations of the ionized fluid and the neutral fluid are essentially decoupled from each other. After the excitation, vortical motions in the ionized fluid oscillate at the Alfv\'en frequency, $k_z\ca$, and with a damping rate proportional to $\nuni$.  These are weakly damped Alfv\'en waves. Conversely, vortical disturbances in the neutral fluid are evanescent in time. There is no driving force for vortical motions in the neutral fluid, and so there is no oscillatory behavior.

\subsubsection{High collision frequency}

Now we turn to the limit of high collision frequency compared to the oscillation frequency. This is the realistic case for many astrophysical applications and deserves special attention. We compute the limit of Equations~(\ref{eq:finalgi}) and (\ref{eq:finalgn}) when $\nuni \gg \omega_{\rm R}$. We obtain
\begin{eqnarray}
 \gammai(t) & \approx & \frac{ \Gamma_{0, \rm i} + \chi \Gamma_{0, \rm n}}{1+\chi} \cos\left( \frac{k_z\ca}{\sqrt{1+\chi}} t \right) \nonumber \\ &\times& \exp\left(  -\frac{\chi}{2\left( 1+\chi \right)^2 } \frac{k_z^2 c_{\rm A}^2 }{\nuni} t \right) \nonumber \\
 &+&  \frac{\chi \left( \Gamma_{0, \rm i} - \Gamma_{0, \rm n} \right) }{1+\chi}\exp \left[ - \left( 1 + \chi \right) \nuin t \right], \label{eq:gilimit} \\
\gamman(t) & \approx & \frac{ \Gamma_{0, \rm i} + \chi \Gamma_{0, \rm n}}{1+\chi}\cos\left( \frac{k_z\ca}{\sqrt{1+\chi}} t \right) \nonumber \\ &\times& \exp\left(  -\frac{\chi}{2\left( 1+\chi \right)^2 } \frac{k_z^2 c_{\rm A}^2 }{\nuni} t \right) \nonumber \\
 &-&  \frac{\Gamma_{0, \rm i} - \Gamma_{0, \rm n}  }{1+\chi}\exp \left[ - \left( 1 + \chi \right) \nuin t \right]. \label{eq:gnlimit}
\end{eqnarray}

By comparing Equations~(\ref{eq:gilimit}) and (\ref{eq:gnlimit}) with the equivalent solutions in the absence of magnetic field (Equations~(\ref{eq:solb0i}) and (\ref{eq:solb0n})) we find that in both cases there is a relaxation phase whose time scale is determined by the collision frequency. In the absence of magnetic field (Equations~(\ref{eq:solb0i}) and (\ref{eq:solb0n})) the vorticity perturbations remain constant after the relaxation phase, while in the presence of magnetic field (Equations~(\ref{eq:gilimit})--(\ref{eq:gnlimit})) the perturbations of  both ions and neutrals oscillate together as a single fluid at the modified Alfv\'en frequency  $k_z \ca/\sqrt{1+\chi}$. The oscillations are exponentially damped in time.

  We define $\Delta(t)$ as the difference of $\gammai(t)$ and $\gamman(t)$ computed from Equations~(\ref{eq:gilimit}) and (\ref{eq:gnlimit}), namely
\begin{equation}
  \Delta(t) \equiv \gammai(t) - \gamman(t) \approx \left( \Gamma_{0, \rm i} -\Gamma_{0, \rm n}  \right) \exp\left(- \frac{t}{\tau_{\rm rel}}\right),
\end{equation}
with $\tau_{\rm rel}$ the relaxation time scale given by
\begin{equation}
\tau_{\rm rel} = \frac{1}{\left(1+\chi\right)\nuni} = \frac{1}{\nuin + \nuni} . \label{eq:timerel}
\end{equation}
The function $\Delta(t)$ informs us about the relaxation phase. The function $\Delta(t) \to 0$ when $ t \gg \tau_{\rm rel}$. Importantly, $\Delta(t) = 0$  when $\Gamma_{0, \rm i} =\Gamma_{0, \rm n} $, i.e., there is no relaxation phase when the initial disturbance perturbs ions and neutrals in the same way. In such a case, both ions and neutrals oscillate together with amplitude equal to the initial condition. When $\Gamma_{0, \rm i} \neq \Gamma_{0, \rm n} $ the amplitude of the Alfv\'enic oscillations is the weighted average of the initial conditions (Equation~(\ref{eq:amplitude})). The weights correspond to the densities of each fluid.

\subsection{Numerical solution}

Here we use the numerical code MolMHD to evolve in time the  basic Equations~(\ref{eq:momlinion})--(\ref{eq:presslin}) in a fully numerical way. The numerical code \citep[see][for details about the scheme]{bona2009} uses the method of lines for the discretization of the variables, and the time and space variables are treated
separately. For the temporal part, a 4th order Runge-Kutta method is used, whereas for the space discretization a finite difference
scheme with a 4th order centered stencil is used. For a given spatial resolution, the time step is selected so as to satisfy
the Courant condition.

The MolMHD code evolves in time the plasma and magnetic field perturbations after an initial condition. In the present application the perturbations are assumed invariant in the $x$- and $y$-directions, so that perturbations only depend on the $z$-direction and Alfv\'en waves are strictly polarized in the $y$-direction.  The only non-zero perturbations are the components of velocity and magnetic field in the $y$-direction. The numerical integration of Equations~(\ref{eq:momlinion})--(\ref{eq:presslin}) is done in the interval $z\in[-10 L, 10 L]$, where $L$ is an arbitrary length scale. We use a uniform grid with 1001 grid points. Since we study standing waves the boundary conditions used at $z=\pm 10L$ are that the velocity perturbations are fixed to zero. The boundary conditions for the other wave perturbations are that their spatial derivatives are set to zero. The initial condition at $t=0$ for the velocity of ions, $v_{{\rm i},y}$, is
\begin{equation}
v_{{\rm i},y}= v_{{\rm i},0} \cos\left( \frac{\pi}{20L}z  \right), \label{eq:vinitial}
\end{equation}
which corresponds to the fundamental standing mode in our domain with dimensionless wavenumber $k_zL = \pi/20$. In the fully ionized case, the dimensionless frequency of the fundamental mode is $\omega L / \ca = \pi/20 \approx 0.157$. Since we are dealing with linear perturbations we express the velocity perturbation in arbitrary units and set $v_{{\rm i},0} = 1$.

\begin{figure*}
	\centering
	\includegraphics[width=.99\columnwidth]{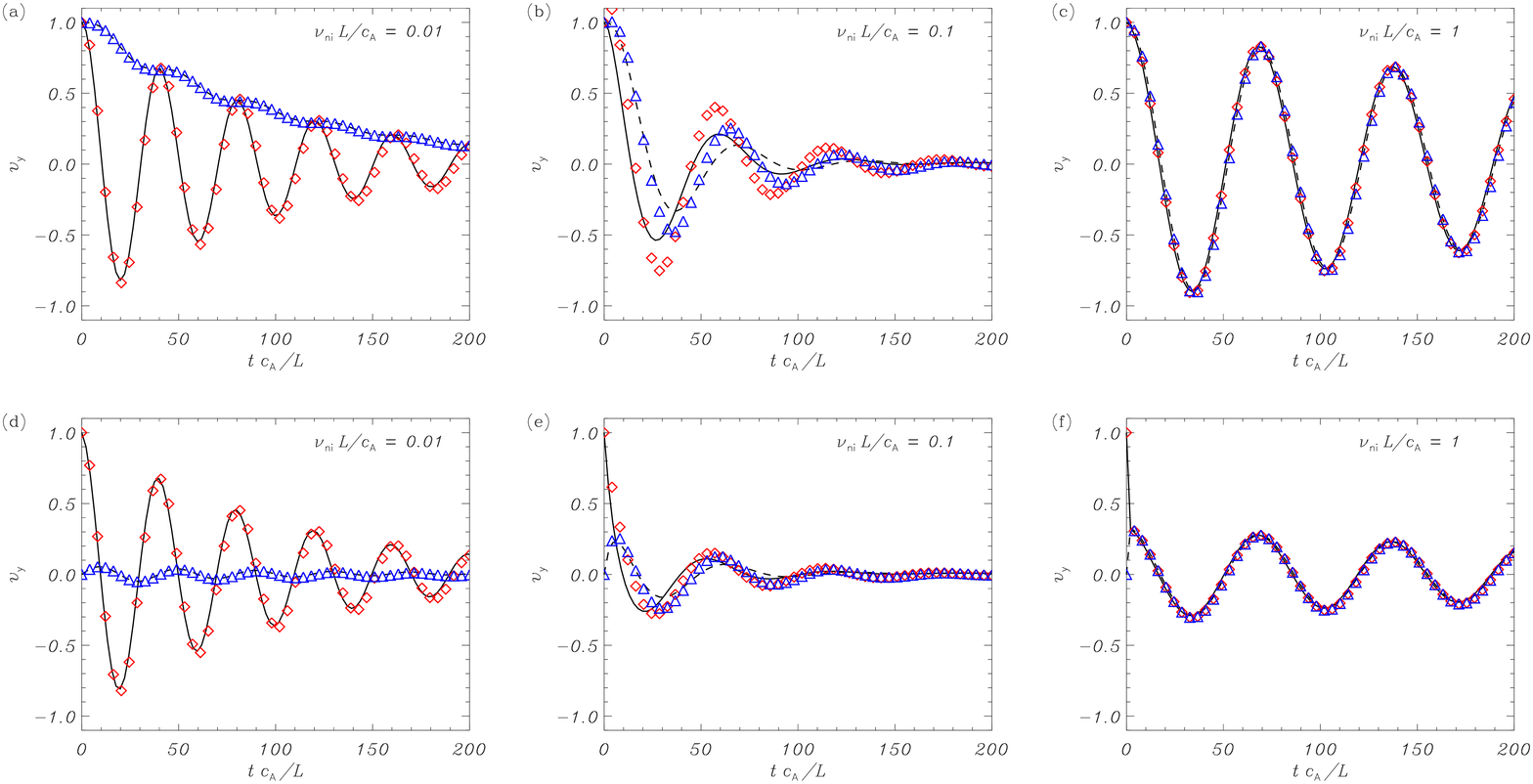}
	\caption{Results of the initial value problem. Velocity perturbation (in arbitrary units) of ions (solid line) and neutrals (dashed line) at $z=0$ as functions of time computed numerically with the MolMHD code for $\nuni L/\ca =$~0.01, 0.1, and 1.  The symbols $\Diamond$ and $\triangle$ are the corresponding analytic results (Equations~(\ref{eq:finalgi}) and (\ref{eq:finalgn})). In panels (a), (b), and (c) we have used the same initial condition for the velocity of both ions and neutrals, while in panels (d), (e), and (f) neutrals are initially at rest. We have used $\chi=2$ in all cases. For reference, the dimensionless ideal Alfv\'en frequency is $\omega L / \ca = \pi/20 \approx 0.157$.}
	\label{fig:mol}
\end{figure*}

Figure~\ref{fig:mol} shows the velocity perturbation  of ions and neutrals  at $z=0$ computed numerically with the MolMHD code for various values of the neutral-ion collision frequency. We compare the numerical results with the analytic ones (Equations~(\ref{eq:finalgi}) and (\ref{eq:finalgn})).  The results in the top row of Figure~\ref{fig:mol} (panels a, b, and c) are obtained when the velocity of neutrals at $t=0$ is the same as that of ions (Equation~(\ref{eq:vinitial})), while in the bottom row of Figure~\ref{fig:mol} (panels d, e, and f) neutrals are initially at rest. In the following paragraphs we discuss the results displayed in Figure~\ref{fig:mol} and relate the time-dependent solutions with the normal modes investigated in Section~\ref{sec:normal}.

(i) In Figure~\ref{fig:mol}(a,d) we use $\nuni L/\ca =0.01$ so that the collision frequency is an order of magnitude lower than the ideal Alfv\'en frequency. Ions and neutrals behave almost independently. Regardless the initial condition for neutrals, ions oscillate at the Alfv\'en frequency $k_z \ca$ and are damped due to collisions. The oscillations of the ionized fluid act as a periodic forcing on neutrals, although the oscillations generated in the neutral fluid are of much lower amplitude compared to those in the ionized fluid. The dominant behavior in neutrals when the initial perturbation is nonzero is the exponential decay of the initial perturbation. This behavior is consistent with Equations~(\ref{eq:gilow}) and (\ref{eq:gnlow}). In relation with the normal modes of Section~\ref{sec:normal}, we see that when the collision frequency is low the oscillatory mode is mostly excited in the ionized fluid, while the evanescent mode dominates the neutral fluid behavior. Hence we can understand the physical reason for the existence of the evanescent mode as the way in which vorticity perturbations in the neutral fluid decay in time due to collisions  \citep[see also][]{zaqarashvili2011a}.

(ii) In Figure~\ref{fig:mol}(b,e) we increase the collision frequency to $\nuni L/\ca =0.1$. The collision frequency and the ideal Alfv\'en frequency are of the same order of magnitude. Now both ions and neutrals display strongly damped oscillations and it is not possible to relate the behavior of each fluid with one particular normal mode. Both oscillatory and evanescent  normal modes are excited in both fluids and the observed behavior is the result of the joint effect of both oscillatory and evanescent modes. We notice that in Figure~\ref{fig:mol}(b,e) the analytic solution of the time-dependent problem does not exactly follow the full numerical result.  The source of the discrepancy is in the approximate expressions of $\omega_{\rm R}$, $\omega_{\rm I}$, and $\epsilon$ (Equations~(\ref{eq:wr})--(\ref{eq:gamma})) used in the analytic solution of the initial-value problem (Equations~(\ref{eq:finalgi}) and (\ref{eq:finalgn})). These approximate expressions were obtained in the case of weak damping, which obviously does not apply to the situation of Figure~\ref{fig:mol}(b,e). However, it is remarkable that the analytic solution still captures the overall behavior correctly even beyond the weak damping limit. It is worth noting that instead of using the approximate values of $\omega_{\rm R}$, $\omega_{\rm I}$, and $\epsilon$ we could use their exact values obtained by numerically solving the dispersion relation. In such a case the agreement between analytic and numerical results is excellent (not shown here for simplicity).

\begin{figure}
	\centering
	\includegraphics[width=.65\columnwidth]{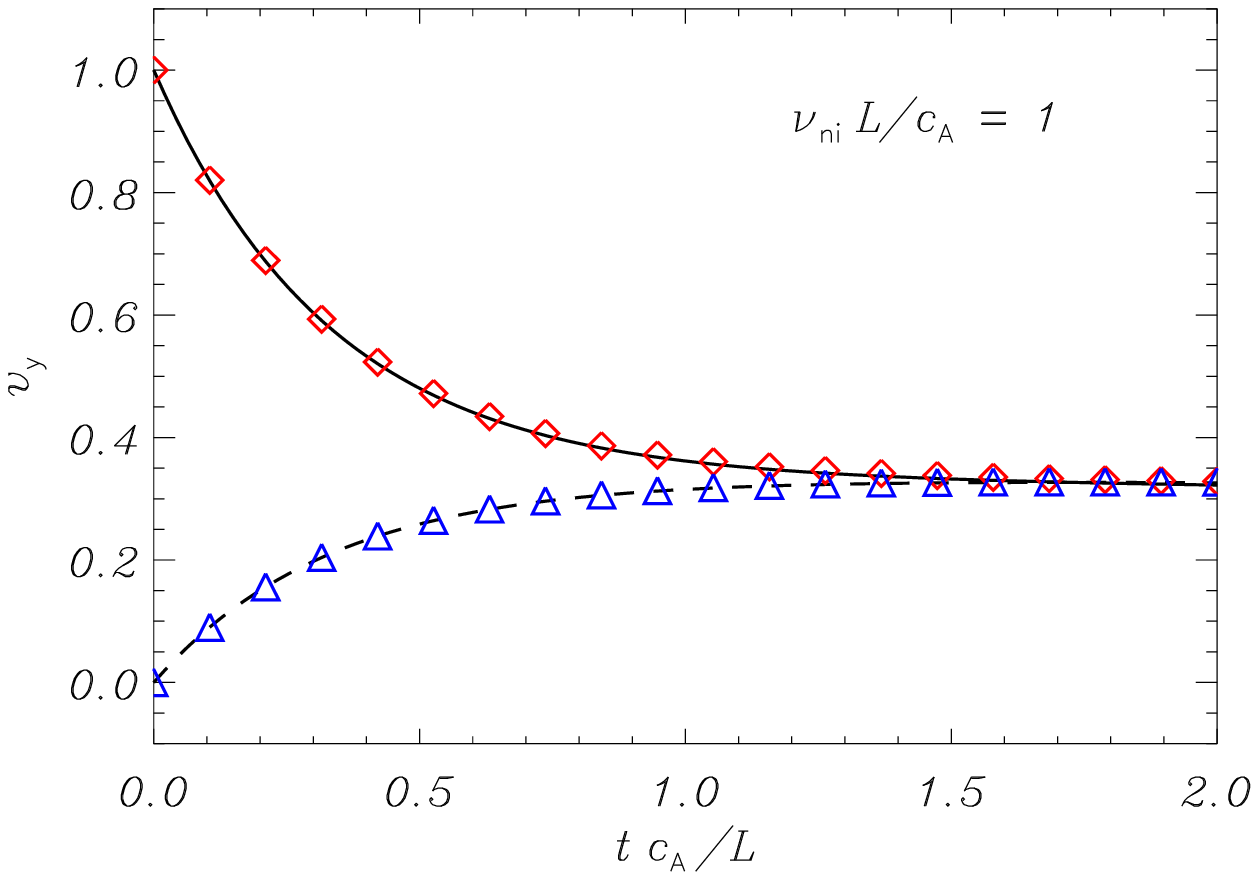}
	\caption{Same as Figure~\ref{fig:mol}(f) but near $t=0$.}
	\label{fig:mol2}
\end{figure}

(iii) Finally we use $\nuni L/\ca =1$ in Figure~\ref{fig:mol}(c,f) so that the collision frequency is an order of magnitude higher than the ideal Alfv\'en frequency. We find that ions and neutral behave as a single fluid and oscillate together at the modified Alfv\'en frequency $k_z \ca/\sqrt{1+\chi}$. Again, both numerical and analytic results are in excellent agreement. The common amplitude of the $y$-components of velocity of ions and neutrals, namely $\hat{v}_{y}$, is
\begin{equation}
\hat{v}_{y} =  \frac{\rhoi v_{{\rm i},0} + \rhon v_{{\rm n},0}}{\rhoi + \rhon}. \label{eq:relvel}
\end{equation} 
The oscillations are weakly damped. When the initial velocity of neutrals is the same as that of ions (Figure~\ref{fig:mol}(c)) both fluids oscillate together from $t=0$. The evanescent mode is not excited when $v_{{\rm i},0} = v_{{\rm n},0}$. Conversely, when neutrals are initially at rest (Figure~\ref{fig:mol}(f)), the single-fluid behavior begins after a short relaxation phase. Although very brief, the relaxation phase can be seen in Figure~\ref{fig:mol}(f) near $t=0$ as sudden decrease of $v_{{\rm i},y}$ and increase of $v_{{\rm n},y}$. Figure~\ref{fig:mol2} shows  the same results displayed in Figure~\ref{fig:mol}(f) but near $t=0$ in order to observe the relaxation phase in detail. The behavior of the velocity perturbations is governed by the evanescent normal mode during the relaxation phase and by the oscillatory normal mode afterwards. This is consistent with Equations~(\ref{eq:gilimit}) and (\ref{eq:gnlimit}). In this case and since $v_{{\rm n},0} = 0$, the amplitude of the joint oscillations after the relaxation phase (Equation~(\ref{eq:relvel})) is
\begin{equation}
\hat{v}_{y} =  \frac{\rhoi }{\rhoi + \rhon} v_{{\rm i},0} =  \frac{1 }{1 + \chi} v_{{\rm i},0}.
\end{equation} 
Hence, the larger the ionization ratio $\chi$, the lower the amplitude.

\section{APPLICATION TO THE LOW SOLAR ATMOSPHERE}
\label{sec:app}

In this Section we perform an application of the previous theoretical analysis to a particular astrophysical plasma, namely the solar atmosphere. We specialize in solar atmospheric plasma because of two main reasons.  The plasma in the coolest parts of the solar atmosphere, i.e., the photosphere and the low chromosphere,  is weakly ionized. In addition, recent observations have shown the ubiquitous presence of Alfv\'enic waves in the solar atmosphere \citep[e.g..,][]{depontieu2007,tomczyk2007,mcintosh2011,okamoto2011,depontieu2012}.  It is therefore interesting to apply the theory developed in the present paper to the solar case. Many works have focused on the damping of chromospheric waves due to ion-neutral collisions and its importance for plasma heating \citep[see, e.g.,][among others]{haerendel1992,depontieu2001,khodachenko2004,leake2005,soler2012,zaqarashvili2013}. In the present application we do not discuss damping but investigate other effects on Alfv\'en waves caused by partial ionization, namely the presence of cut-off wavelengths of standing waves and the modification of the effective Alfv\'en velocity.  For results about wave damping, the reader is referred to the papers cited above.

\begin{figure}
	\centering
	\includegraphics[width=.65\columnwidth]{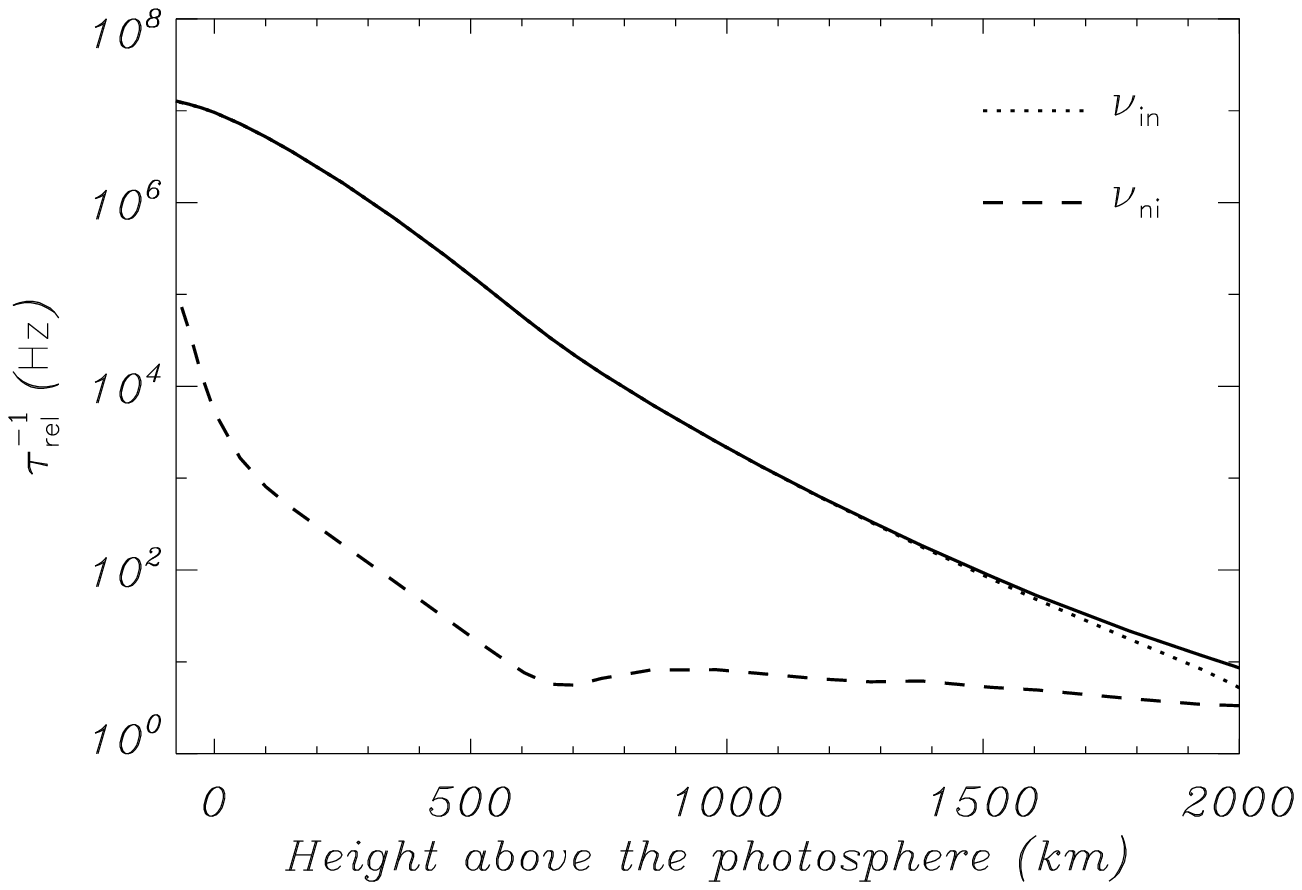}
	\caption{Inverse of the relaxation time of the ion-neutral coupling (solid line) in the low solar atmosphere according to the VALC model. The ion-neutral (dotted line) and neutral-ion (dashed line) collision frequencies are shown for comparison.}
	\label{fig:reltime}
\end{figure}

We consider a simplified model for the solar atmospheric plasma. The variation of physical parameters with height from the photosphere to the base of the corona is taken from the quiet sun model C of \citet{vernazza1981}, hereafter VALC model. To compute the friction coefficient, $\ain$, we ignore the influence of heavier species and consider only hydrogen. Hence, the expression of  $\ain$ is given in Equation~(\ref{eq:ainhydrogen}), where we take $\sigma_{\rm in}\approx 10^{-20}$~m$^2$ according to the estimation by \citet{zaqarashvili2013} for the case of direct elastic collisions \citep[see also][]{brag}. Figure~\ref{fig:reltime} shows the inverse of the relaxation time of the ion-neutral coupling, $\tau_{\rm rel}^{-1}$ (Equation~(\ref{eq:relvel})), as a function of height in the low  solar atmosphere. The very large values of $\tau_{\rm rel}^{-1}$, i.e., very short $\tau_{\rm rel}$, point out that in the solar atmosphere ions and neutrals are very efficiently coupled and, for practical purposes, they can be considered as a single fluid. The single-fluid approach is usually followed in studies focused on wave damping (see the references in the previous paragraph). For comparison, we also plot in   Figure~\ref{fig:reltime} the ion-neutral, $\nuin$, and neutral-ion, $\nuni$, collision frequencies \citep[see also similar plots in][]{depontieu2001}. We find that the value of $\tau_{\rm rel}$ is mainly determined by $\nuin$. Note that in Figure~\ref{fig:reltime} the solid and dotted lines are superimposed except at large heights.

\begin{figure*}[ht]
	\centering
	\includegraphics[width=.65\columnwidth]{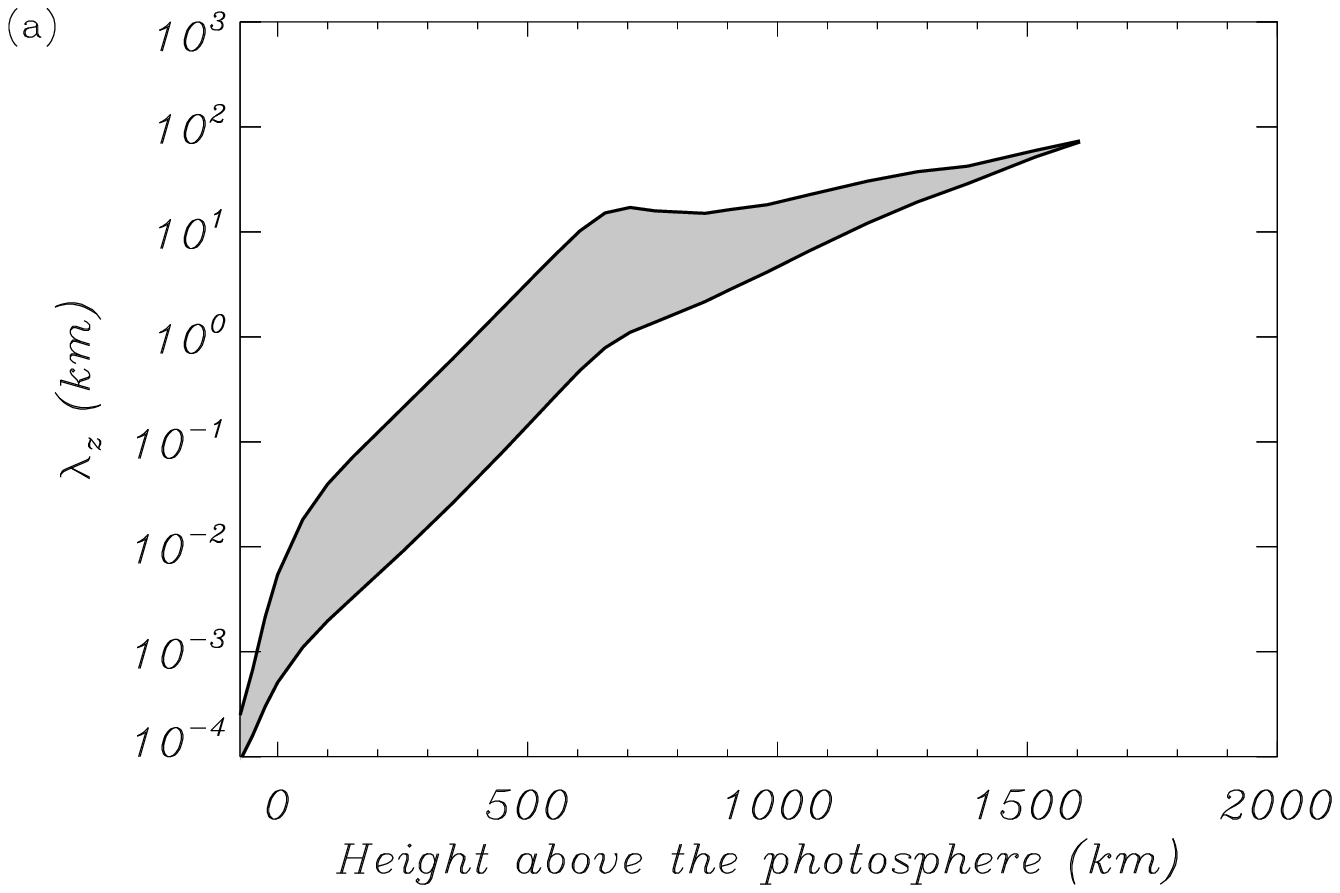}
   \includegraphics[width=.65\columnwidth]{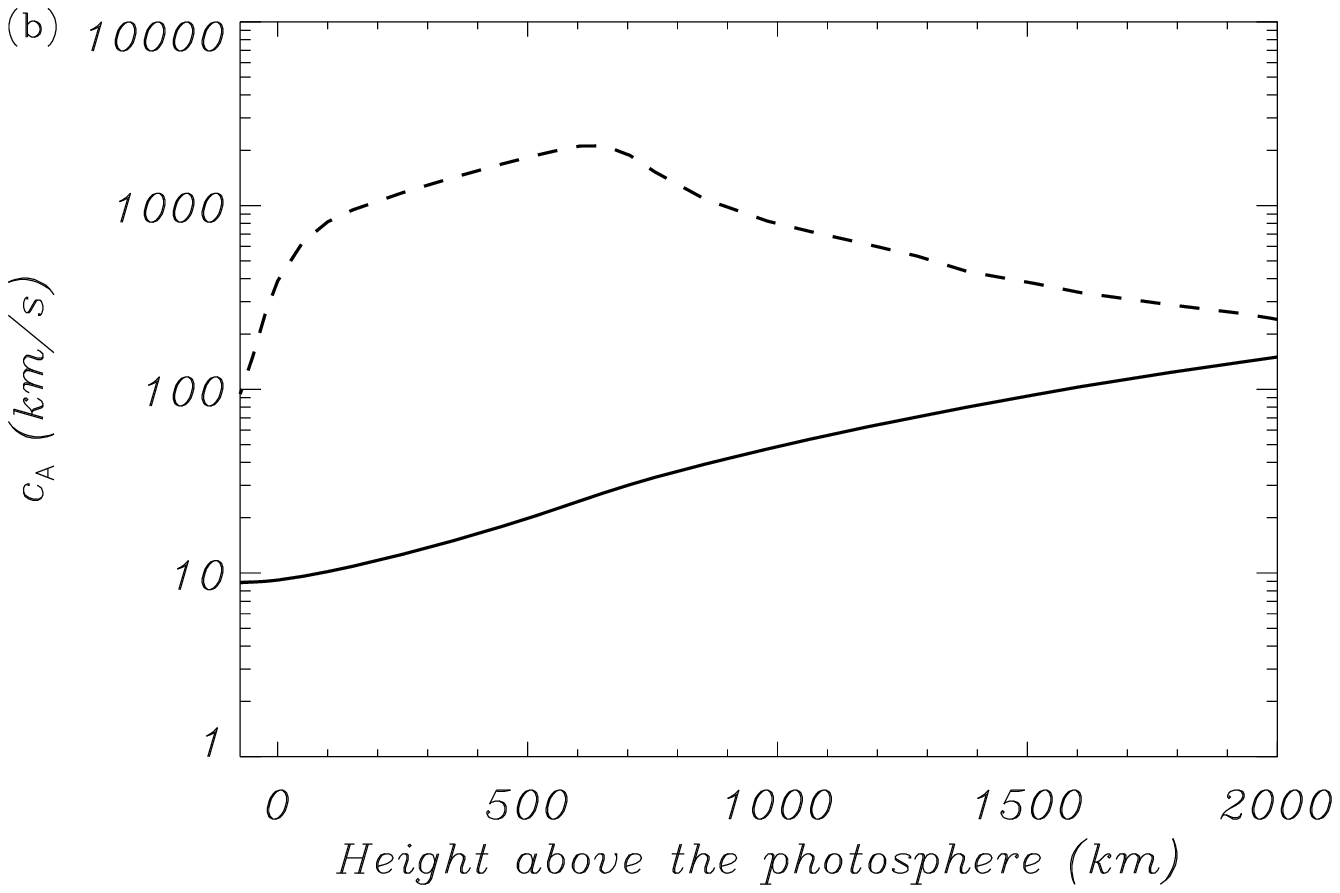}
	\caption{(a) Cut-off region of wavelengths of standing Alfv\'en waves as a function of height in the low solar atmosphere. The shaded area denotes the region where oscillatory standing modes are not possible. (b) Effective Alfv\'en velocity (solid line) as a function of height for a wave frequency of 22 mHz. The  Alfv\'en velocity computed using the ion density only (dashed line) is shown for comparison.}
	\label{fig:valc}
\end{figure*}

Concerning the magnetic field strength, we consider the model by \citet{leake2006} of a vertical chromospheric magnetic flux tube expanding  with height, whose form is given by
\begin{equation}
B = B_{\rm ph} \left( \frac{\rho}{\rho_{\rm ph}} \right)^\beta,
\end{equation}
where $\rho = \rhoi + \rhon$ is the total density, $B_{\rm ph}$ and $\rho_{\rm ph}$ are the magnetic field strength and the total density, respectively, at the photospheric level, and $\beta=0.3$ is an empirical exponent. We use $\rho_{\rm ph} = 2.74\times10^{-4}$~kg~m$^{-3}$ from the VALC model and $B_{\rm ph} = 1.5$~kG. The dependence of $\rho$ with height is determined by the VALC model. With this choice of parameters the magnetic field strength decreases with height so that $B\approx 100$~G at 1,000~km  and  $B\approx 20$~G at 2,000~km above the photosphere.

The physical parameters in the simplified model of the low solar atmosphere used here depend on the vertical direction, $z$, while in the theoretical analysis of the previous Sections all the parameters are taken constant in $z$. However, it is possible to apply the expressions derived before for a homogeneous plasma to the present stratified case if the wavelength in the $z$-direction,  $\lambda_z = 2\pi/k_z$, is much shorter than the length scale of the variation of the effective Alfv\'en velocity. If such a restriction is fulfilled, we can perform a local analysis and use  the physical parameters at a given height in the expressions derived for homogeneous plasma. We follow this approach.

Another assumption made in this Section is that the amplitudes of the waves are small enough for the linear analysis to remain valid. This is a reasonable assumption for chromospheric waves. A parameter that can quantify nonlinearity of the waves is the ratio of the wave velocity amplitude to the local Alfv\'en velocity. The median of the velocity amplitudes of the chromospheric waves detected by \citet{okamoto2011} is 7.4~km~s$^{-1}$. This value is, at least, an order of magnitude lower than the expected value of the Alfv\'en velocity in the chromosphere.

We start by computing the cut-off region of wavenumbers, $k_z$, of standing waves (Equation~(\ref{eq:rmasmenos})) as a function of height. Instead of $k_z$, we show in Figure~\ref{fig:valc}(a) the corresponding wavelength, $\lambda_z$. The shaded area in Figure~\ref{fig:valc}(a) indicates the range of wavelengths where oscillatory standing modes are not possible. At low heights above the photosphere the cut-off wavelengths are very short. The cut-off wavelengths increase with height and become of the order of a few kilometers at heights between 500~km and 1,500~km, approximately. Then the cut-off region disappears at a height of 1,600~km above the photosphere, approximately, because the ionization ratio, $\chi$, decreases with height as the plasma becomes more and more ionized and the threshold value $\chi = 8$ is reached at that height.

Next we turn to the computation of the effective Alfv\'en velocity (Equation~(\ref{eq:effva})). Since this quantity depends on the wave frequency we use a frequency of 22 mHz, which corresponds to the dominant frequency of the chromospheric Alfv\'enic waves detected by \citet{okamoto2011}. We display  in Figure~\ref{fig:valc}(b) the real part of the effective Alfv\'en velocity as a function of height. In Figure~\ref{fig:valc}(b) we also show the corresponding Alfv\'en velocity computed using its classic definition that depends on the ion density only. We find that in the low chromosphere neutral-ion collisions are crucial to decrease the effective Alfv\'en velocity between one and two orders of magnitude with respect to the expected value taking into account the ion density only.

Now we can check the assumption that $\lambda_z$ is much shorter than the length scale of the variation of the effective Alfv\'en velocity. On the one hand from Figure~\ref{fig:valc}(b) we get that the effective Alfv\'en velocity varies about an order of magnitude in 2,000~km, approximately. On the other hand  the largest cut-off wavelengths displayed in Figure~\ref{fig:valc}(a) are of the order of a few tens of kilometers. Hence the use of a local analysis in the present application is justified.

\section{Implication for energy estimates}
\label{sec:energy}

In the solar context, the dissipation of Alfv\'enic waves may play an important role for the heating of the atmospheric plasma \citep[see, e.g.,][]{depontieu2007,mcintosh2011}. The implications of partial ionization for the calculations of energy carried by Alfv\'en waves were discussed by, e.g., \citet{vranjes2008} and \citet{tsap2011}. Both papers give expressions for the energy flux of Alfv\'en waves in a partially ionized plasmas. However the equations provided by \citet{vranjes2008} and \citet{tsap2011} are different. The expression of \citet{vranjes2008} includes the factor  $\left( \rhoi / \rhon \right)^2$, which is absent from the expression of \citet{tsap2011}. Hence the conclusions of these two papers regarding the impact of partial ionization are in apparent contradiction. On the one hand, \citet{vranjes2008} explained that due to the factor $\left( \rhoi / \rhon \right)^2$ the energy flux in weakly ionized  plasmas becomes orders of magnitude smaller compared to the fully ionized case. On the other hand, since \citet{tsap2011} lacked that factor, they argued that the energy flux   is independent of the ionization degree and  obtained the same expression as for fully ionized plasmas. Here we shall try to solve this apparent contradiction between the results of \citet{vranjes2008} and \citet{tsap2011}.

For application to the solar atmosphere we take the limit of high collision frequency. We choose a reference frame in which $k_y = 0$. After the relaxation phase, i.e., for $ t \gg \tau_{\rm rel}$, and neglecting  the weak damping due to ion-neutral collisions, the $y$-components of velocity of ions and neutrals oscillate together with frequency,  $k_z \ca/\sqrt{1+\chi}$, and amplitude, $\hat{v}_{y}$, given in Equation~(\ref{eq:relvel}). The energy flux along magnetic field lines, $ S_z  $, calculated using the Poynting vector \citep[see, e.g.,][]{walker2005} is
\begin{equation}
S_z= \frac{1}{2 \mu} \hat{E}_{x} \hat{b}_{y},
\end{equation} 
where $\hat{E}_{x}$ and $\hat{b}_{y}$ denote the amplitudes of the $x$-component of the electric field and the $y$-component of the magnetic field perturbation. The electric field is
\begin{equation}
{\bf E} = - {\bf v}_{\rm i} \times {\bf B},
\end{equation}
so that $\hat{E}_{x}\sim\hat{v}_{y} B$. On the other hand, the magnetic field perturbation is governed by Equation~(\ref{eq:inductionlin}), from where we get $\hat{b}_{y}\sim \hat{v}_{y} B \sqrt{1+\chi}/ \ca$.  Hence the energy flux is
\begin{equation}
 S_z = \frac{\sqrt{1+\chi}}{\ca}\frac{B^2}{2\mu} \hat{v}_{y}^2= \frac{B}{2\sqrt{\mu}} \sqrt{\rho} \, \hat{v}_{y}^2, \label{eq:energy1}
\end{equation}
with $\rho = \rhoi+\rhon$ the total plasma density. Consistently we find the classical expression of the energy flux of Alfv\'en waves \citep[see, e.g.,][]{walker2005}, with the only difference that the total plasma density replaces ion density and that the velocity amplitude after the relaxation phase,  $\hat{v}_{y}$, is used. Since the relaxation phase is extremely short for realistic collision frequencies (Figure~\ref{fig:reltime}),  $\hat{v}_{y}$ is the velocity amplitude that an observer would actually measure. In a fully ionized plasma, $\hat{v}_{y} = v_{{\rm i},0}$ so that
\begin{equation}
 S_z =  \frac{B}{2\sqrt{\mu}} \sqrt{\rho} \, v_{{\rm i},0}^2. \label{eq:energy12}
\end{equation}

Up to here there is no discrepancy between the analysis of \citet{vranjes2008} and \citet{tsap2011}. The discrepancy arises when the energy flux is written in terms of the initial velocity amplitudes of ions and neutrals. We insert Equation~(\ref{eq:relvel})  in Equation~(\ref{eq:energy1}) and rewrite the energy flux in terms of the initial velocity amplitudes, namely
\begin{equation}
 S_z  =  \frac{B}{2\sqrt{\mu}} \sqrt{\rho} \left( \frac{\rhoi v_{{\rm i},0} + \rhon v_{{\rm n},0}}{\rhoi + \rhon} \right)^2,
\end{equation}
In weakly ionized plasmas $\rhon \gg \rhoi$ and the expression simplifies to
\begin{equation}
 S_z  \approx  \frac{B}{2\sqrt{\mu}} \sqrt{\rho} \left( \frac{\rhoi}{\rhon} v_{{\rm i},0} + v_{{\rm n},0}\right)^2. \label{eq:energy2}
\end{equation} 
In the analysis of \citet{vranjes2008} the  initial disturbance is localized in the ionized fluid only. So \citet{vranjes2008} took $v_{{\rm n},0}=0$. In this case Equation~(\ref{eq:energy2}) becomes
\begin{equation}
 S_z \approx  \frac{B}{2\sqrt{\mu}} \sqrt{\rho} \left( \frac{\rhoi}{ \rhon} \right)^2v_{{\rm i},0}^2. \label{eq:vran}
\end{equation} 
Due to the presence of the factor $\left( \rhoi / \rhon \right)^2$ the energy flux computed by \citet{vranjes2008} in a weakly ionized plasma is much lower than the corresponding value in a fully ionized plasma (Equation~(\ref{eq:energy12})). The physical reason is that significant portion of the wave energy is used to set into motion the neutral fluid, which is initially at rest in \citet{vranjes2008}. 

On the contrary, \citet{tsap2011} imposed $\hat{v}_{y} = v_{{\rm i},0}$, which can be  satisfied only if $v_{{\rm i},0} = v_{{\rm n},0}$ according to Equation~(\ref{eq:relvel}). This means that \citet{tsap2011} implicitly assumed that both ions and neutrals are initially perturbed with the same velocity, although this condition is not explicitly stated in the paper.  The energy flux is then
\begin{equation}
S_z  \approx  \frac{B}{2\sqrt{\mu}} \sqrt{\rho} \, v_{{\rm i},0}^2. \label{eq:tsap}
\end{equation}
The expression found by  \citet{tsap2011} is the same as Equation~(\ref{eq:energy12}) for a fully ionized plasma. This is so because in \citet{tsap2011} neutrals have initially the same velocity as ions and   no wave energy needs to be used in setting neutrals into motion. \citet{tsap2011} claimed that the expression of \citet{vranjes2008} is incorrect. Here we clearly see that the discrepancy between Equations~(\ref{eq:vran}) and (\ref{eq:tsap}) is caused by a different choice of the initial conditions and that both expressions are indeed correct. The source of the discrepancy was  expressing the energy flux in terms of the initial velocity of ions. This is a velocity definitely not easy to measure observationally. Instead, the velocity that an observer can measure is the velocity amplitude after the relaxation phase,  $\hat{v}_{y}$. Then, Equation~(\ref{eq:energy1}) should be used to estimate the energy flux.

Finally we recall that the present analysis, and so Equation~(\ref{eq:energy1}), is valid in the case of Alfv\'en waves propagating in a uniform medium so that the energy flux is uniform. As commented in Section~\ref{sec:app}, the application of Equation~(\ref{eq:energy1}) to the case of Alfv\'enic waves in the solar atmosphere must be done with caution since neither the density nor the wave velocity amplitude are uniform in the solar plasma \citep[see][]{goossens2013}.

\section{Conclusions}
\label{sec:con}

In this paper we have studied theoretically  Alfv\'en waves in a plasma composed of an ion-electron fluid and a  neutral fluid interacting by means of neutral-ion collisions. This configuration is relevant in many astrophysical and laboratory plasmas. To keep our investigation as general as possible we have taken the neutral-ion collision frequency  and the ionization degree as  free parameters. 

First we have performed a normal mode analysis and have derived the dispersion relation of linear Alfv\'en waves. The dispersion relation agrees with the equations previously found by, e.g., \citet{kulsrud1969,pudritz1990,martin1997,kumar2003,zaqarashvili2011a,mouschovias2011}. As in previous studies, we find that Alfv\'en waves are damped by neutral-ion collisions. The damping is most efficient when the wave frequency and the collision frequency are of the same order of magnitude. This conclusion applies to both standing and propagating waves. The effective Alfv\'en velocity of the plasma is modified by collisions, so that for high collision frequencies compared to the wave frequency the effective Alfv\'en velocity depends on the total density of the plasma and not  on the ion density only \citep[see also][]{kumar2003}. 

In addition, an important result is that when $\chi > 8$, i.e., when the plasma is weakly ionized, there is a range of  wavenumbers (or, equivalently, of wavelengths) for which oscillatory standing Alfv\'en waves are not possible. These cut-off wavenumbers are physical and are caused by a purely two-fluid effect  \citep[see][]{kulsrud1969,pudritz1990,kamaya1998,mouschovias2011}. We investigated the physical reason for the existence of this cut-off region by analyzing the relative importance of the magnetic tension force and the neutral-ion friction force. We found that friction becomes the dominant force in the cut-off region, while tension is more important outside the cut-off interval. Hence, a disturbance in the magnetic field whose wavenumber is within the cut-off range decays due to collisions before the plasma is able to feel the restoring force of magnetic tension. As a consequence, oscillations of the magnetic field are effectively suppressed. The cut-off wavenumbers investigated here do not appear in the single-fluid approximation and are different from the unphysical cut-offs found in the single-fluid approximation \citep{zaqarashvili2012}.

The solution of the initial-value problem is fully consistent with the normal modes analysis, and shows a growing strength of the interaction between ions and neutrals as the collision frequency increases. For high collision frequencies both ions and neutrals behave as a single fluid. Here a `high collision frequency' means that the neutral-ion collision frequency is at least an order of magnitude higher than the wave frequency. This condition is fulfilled in many astrophysical plasmas which makes the single-fluid approximation appropriate in those situations for the computations of  periods/wavelengths and damping times/damping lengths. However, as explained above the single-fluid limit  does not fully capture the details of the interaction between ions and neutrals and, for example, the existence of a range of cut-off wavenumbers is a two-fluid result.

As an example, we have considered Alfv\'en waves in a plasma with physical condition akin to those in the low solar atmosphere. As a matter of fact, due to the large values of the collision frequency this plasma could be studied using the single-fluid approximation instead of the more general two-fluid theory. In that respect, the effective Alfv\'en velocity is found to depend on the total density of the plasma. However, the presence of a certain range of cut-off wavelengths that can constrain the existence of oscillatory standing modes in the chromosphere is  a result absent from those previous studies that use the limit of strong coupling \citep[e.g.,][]{haerendel1992,depontieu1998,khodachenko2004,leake2005}. There are other astrophysical situations in which the ion-neutral coupling is not so strong as in the solar atmosphere and, therefore, a two-fluid treatment is needed. This may be the case of, e.g., protostellar discs \citep[see a table with values of some physical parameters realistic of protostellar discs in][]{malyshkin2011} and molecular clouds \citep[see, e.g.,][]{balsara1996,mouschovias2011}. 

In the present work we have restricted ourselves to Alfv\'en waves. Compressional magnetoacoustic waves have not been investigated. It is expected that both types of MHD waves are simultaneously present in a plasma. Magnetoacoustic waves in a two-fluid plasma will be investigated in a forthcoming work.

\begin{acknowledgements}{}
We acknowledge the support  from MINECO and FEDER Funds through grant AYA2011-22846 and from CAIB through the `grups competitius' scheme and FEDER Funds. JT acknowledges support from MINECO through a Ram\'on y Cajal grant. RS thanks Ramon Oliver for reading the manuscript and for his constructive criticism.
\end{acknowledgements}

\bibliographystyle{apj}
\bibliography{refs}

\begin{appendix}

\section{Expressions of Coefficients}
\label{app}

The expressions of the coefficients of Equations~(\ref{eq:finalgi}) and (\ref{eq:finalgn}) are as follows,
\begin{eqnarray}
A_1 &=& \epsilon\frac{\left(  \epsilon + \nuni \right)\Gamma_{0, \rm i} + \chi \nuni \Gamma_{0, \rm n}}{\omega_{\rm R}^2 + \left( \epsilon - \omega_{\rm I} \right)^2}, \\
A_2 &=& \frac{\left[ \omega_{\rm R}^2 + \omega_{\rm I}^2 - \epsilon \left( \nuni + 2 \omega_{\rm I} \right) \right]\Gamma_{0, \rm i} + \chi \nuni \epsilon \Gamma_{0, \rm n}}{\omega_{\rm R}^2 + \left( \epsilon - \omega_{\rm I} \right)^2}, \\
A_3 &=& \frac{\omega_{\rm R}^2 \left( \epsilon + \nuni + \omega_{\rm I} \right) - \omega_{\rm I}\left( \epsilon - \omega_{\rm I}  \right) \left( \nuni + \omega_{\rm I} \right)}{\omega_{\rm R}^2 + \left( \epsilon - \omega_{\rm I} \right)^2} \Gamma_{0, \rm i}  +  \frac{\chi\nuni \left( \omega_{\rm R}^2 + \omega_{\rm I}^2 - \epsilon \omega_{\rm I} \right)}{\omega_{\rm R} \left[ \omega_{\rm R}^2 + \left( \epsilon - \omega_{\rm I} \right)^2 \right]} \Gamma_{0, \rm n}, \\
C_1 &=& \frac{\nuni\epsilon}{\epsilon + \nuni} \frac{\left(  \epsilon + \nuni \right)\Gamma_{0, \rm i} + \chi \nuni \Gamma_{0, \rm n}}{\omega_{\rm R}^2 + \left( \epsilon - \omega_{\rm I} \right)^2}, \\
C_2 &=& \left( 1 + \frac{\chi\nuni^3}{\left[ \omega_{\rm R}^2 + \left( \nuni + \omega_{\rm I} \right)^2  \right]  \left( \epsilon + \nuni \right)} \right) \Gamma_{0, \rm n}, \\
C_3 &=& - \frac{\epsilon \nuni}{\omega_{\rm R}^2 + \left( \epsilon - \nuni \right)^2}  \Gamma_{0, \rm i}  - \frac{\chi\nuni^2 \left(  \omega_{\rm R}^2 + \omega_{\rm I}^2 + \epsilon\nuni \right)}{\left[ \omega_{\rm R}^2 + \left( \nuni + \omega_{\rm I} \right)^2  \right] \left[  \omega_{\rm R}^2 + \left( \epsilon - \omega_{\rm I} \right)^2\right]}  \Gamma_{0, \rm n}, \\
C_4 &=&  \frac{\nuni\left( \omega_{\rm R}^2 + \omega_{\rm I}^2 - \epsilon \omega_{\rm I} \right)}{\omega_{\rm R}^2 + \left( \epsilon - \nuni \right)^2}  \Gamma_{0, \rm i}  + \frac{\chi\nuni^2 \left[  \omega_{\rm R}^2 \left(  \omega_{\rm I} - \epsilon +\nuni  \right) + \omega_{\rm I} \left( \nuni + \omega_{\rm I} \right) \left( \omega_{\rm I} - \epsilon \right) \right]}{\omega_{\rm R}\left[ \omega_{\rm R}^2 + \left( \nuni + \omega_{\rm I} \right)^2  \right] \left[  \omega_{\rm R}^2 + \left( \epsilon - \omega_{\rm I} \right)^2\right]}  \Gamma_{0, \rm n},
\end{eqnarray}
with the expressions of $\omega_{\rm R}$, $\omega_{\rm I}$, and $\epsilon$  given in Equations~(\ref{eq:wr})--(\ref{eq:gamma}). In the expression of $\omega_{\rm R}$ (Equation~(\ref{eq:wr})) the $+$ sign has to be taken.

\end{appendix}

\end{document}